\documentclass[10pt,conference]{IEEEtran}
\usepackage{cite}
\usepackage{amsmath,amssymb,amsfonts}
\usepackage{algorithm}
\usepackage{algpseudocode}
\usepackage{graphicx}
\usepackage{textcomp}
\usepackage[table]{xcolor}
\usepackage{subcaption}
\usepackage{tabularx}
\usepackage{booktabs}
\usepackage[most]{tcolorbox}
\usepackage{mdframed}
\usepackage{tikz}
\usepackage{quantikz}
\usepackage[hyphens]{url}
\usepackage{fancyhdr}
\usepackage{hyperref}

\usepackage[inline]{enumitem}
\setlist{noitemsep,topsep=0pt,parsep=0pt,partopsep=0pt}

\usepackage{setspace}
\usepackage[margin=5pt,font={stretch=0.9}]{caption}

\newcommand{\hpcayear}{2027}

\newcommand{\hpcasubmissionnumber}{3539}
\title{\textsc{Harvest}: Resource-Aware Quantum Compilation for
Magic State Protocols}

\def\hpcacameraready{} 

\newcommand\hpcaauthors{Jannik Pflieger, Aleksandra Świerkowska, Emmanouil Giortamis, Pramod Bhatotia}
\newcommand\hpcaaffiliation{Technical University of Munich}
\newcommand\hpcaemail{\{firstname\}.\{lastname\}@tum.de}

\providecommand{\Description}[1]{}

\newcommand{\myparagraph}[1]{\noindent{\bf {#1}.}}

\newcommand{\researchquestion}[2][Research question]{%
\begin{tcolorbox}[
    enhanced,
    sharp corners,
    colback=white,
    colframe=black!65,
    boxrule=1.2pt,
    left=0.7em,
right=0.7em,
top=0.8em,
bottom=0.45em,
boxsep=0pt,
before skip=0.4em,
after skip=0.4em,
    title=#1,
    fonttitle=\bfseries,
    coltitle=black,
    attach boxed title to top left={
        xshift=1.7em,
        yshift=-0.75\baselineskip
    },
    boxed title style={
        colback=white,
        colframe=white,
        boxrule=0pt,
        left=0.25em,
        right=0.25em,
        top=0pt,
        bottom=0pt
    }
]
{#2}
\end{tcolorbox}
}

\newcommand{\statementLeftAccent}[3]{%
  \begin{mdframed}[%
    backgroundcolor=#2!5,          
    linecolor=#2!70!black,         
    linewidth=4pt,                 
    topline=false,                 
    rightline=false,               
    bottomline=false,              
    innertopmargin=4pt,
    innerbottommargin=4pt,
    innerleftmargin=12pt,
    skipabove=4pt,
    skipbelow=4pt
  ]%
  {\textbf{#1:} #3}%
  \end{mdframed}%
}

\newcommand{\sysname}[1]{\textsc{Harvest}}

\definecolor{supportgreen}{RGB}{220,240,220}
\definecolor{partialyellow}{RGB}{255,244,204}
\definecolor{unsupportedred}{RGB}{248,220,220}
\definecolor{codegreen}{rgb}{0,0.6,0}

\algrenewcommand{\algorithmiccomment}[1]{\textcolor{codegreen}{// #1}}

\newcommand{\fullsupport}{\cellcolor{supportgreen}\(\checkmark\)}
\newcommand{\partialsupport}{\cellcolor{partialyellow}\(\circ\)}
\newcommand{\nosupport}{\cellcolor{unsupportedred}--}

\author{
  \ifdefined\hpcacameraready
    \IEEEauthorblockN{\hpcaauthors{}}
      \IEEEauthorblockA{
        \hpcaaffiliation{} \\
        \hpcaemail{}
      }
  \else
    \IEEEauthorblockN{\normalsize{HPCA \hpcayear{} Submission
      \textbf{\#\hpcasubmissionnumber{}}} \\
      \IEEEauthorblockA{
        Confidential Draft \\
        Do NOT Distribute!!
      }
    }
  \fi 
}

\fancypagestyle{camerareadyfirstpage}{%
  \fancyhead{}
  
  \fancyhead[C]{
    \ifdefined\aeopen
    \parbox[][12mm][t]{13.5cm}{\hpcayear{} IEEE International Symposium on High-Performance Computer Architecture (HPCA)}    
    \else
      \ifdefined\aereviewed
      \parbox[][12mm][t]{13.5cm}{\hpcayear{} IEEE International Symposium on High-Performance Computer Architecture (HPCA)}
      \else
      \ifdefined\aereproduced
      \parbox[][12mm][t]{13.5cm}{\hpcayear{} IEEE International Symposium on High-Performance Computer Architecture (HPCA)}
      \else
      \parbox[][0mm][t]{13.5cm}{\hpcayear{} IEEE International Symposium on High-Performance Computer Architecture (HPCA)}
    \fi 
    \fi 
    \fi 
    \ifdefined\aeopen 
      \includegraphics[width=12mm,height=12mm]{ae-badges/open-research-objects.pdf}
    \fi 
    \ifdefined\aereviewed
      \includegraphics[width=12mm,height=12mm]{ae-badges/research-objects-reviewed.pdf}
    \fi 
    \ifdefined\aereproduced
      \includegraphics[width=12mm,height=12mm]{ae-badges/results-reproduced.pdf}
    \fi
  }
  \fancyfoot[C]{}
}
\begin{document}
\maketitle

\ifdefined\hpcacameraready 
  \pagestyle{empty}
\else
  \thispagestyle{plain}
  \pagestyle{plain}
\fi

\newcommand{\hpcaheight}{0mm}
\ifdefined\eaopen
\renewcommand{\hpcaheight}{12mm}
\fi

\begin{abstract}
Fault-tolerant quantum processors based on topological codes execute programs through lattice surgery, where operations must be mapped, routed, and supplied with magic states across a 2D grid of physical patches. Non-Clifford operations require these magic states, produced either by distillation factories or by cultivation, each trading footprint against preparation latency, and delivering a magic state to the data patches that consume it requires routing through the same shared layout as every other operation. Yet placement, routing, scheduling, and magic-state supply cannot be optimized in isolation: two operations with no circuit-level dependency can still contend for the same ports, routes, or magic-state terminals once placed, so a compiler that decouples instruction scheduling from magic-state generation, or hard-codes a single generation protocol, is forced to trade execution time against layout footprint instead of co-optimizing both across protocols.

We present \sysname{}, a resource-aware compilation approach for lattice-surgery that co-optimizes magic-state consumption with circuit-aware placement and congestion-aware routing under a protocol-agnostic resource model, then reclaims unused layout footprint after scheduling. Across standard benchmark suites (QAOA, QFT, QASMBench), \sysname{} achieves an average speedup of $4.83\times$ (up to $17.8\times$) over sequential execution, improves schedule length by up to $1.35\times$ through circuit-aware placement, and reclaims up to $72.0\%$ of unused magic-state patches and $33.9\%$ of unused routing patches.

\end{abstract}

\section{Introduction}
\label{sec:introduction}

Fault-tolerant quantum computing (FTQC) requires quantum error correction (QEC) to keep logical qubits protected from noise at scale~\cite{Shor1995,AharonovBenOr2008,gottesman1998theory,Gottesman2009QECReview,terhal2015quantum,Nielsen_Chuang_2010}. In practice, this protection is realized by \emph{topological} codes, which encode each logical qubit in a two-dimensional patch of physical qubits and correct errors through local, nearest-neighbor operations~\cite{DennisEtAl2002,Fowler_2012}. Because each logical qubit occupies its own patch, a fault-tolerant operation between two logical qubits means acting on two patches at once; these codes realize such operations through \emph{lattice surgery}, temporarily merging and splitting neighboring patches to perform joint logical measurements~\cite{HorsmanEtAl2012,litinski_game_2019}. Lattice surgery alone, however, is not sufficient for universal computation, since it only implements Clifford operations.

Non-Clifford operations, such as the \(T\) gate, instead consume a \emph{magic state}, a separately prepared resource state~\cite{Bravyi_2005,Bravyi_2012}. Magic states can be produced in diverse ways: distillation purifies many noisy copies into fewer high-fidelity ones using large dedicated factories, while cultivation grows a state directly, trading a smaller footprint for extra preparation latency and different placement constraints; neither approach dominates the other~\cite{gidney2024magicstatecultivationgrowing,hofmeyr_scheduling_2026}.

This raises a natural question: \emph{how does a magic state, prepared somewhere on the layout, actually reach the data patches that need to consume it?} A magic state and the data patches of the operation that consumes it are, in general, not adjacent, so for every non-Clifford operation in the program, the compiler must connect them through a path of intermediate routing patches~\cite{paler2020opensurgerytopologicalassemblies,paler2019surfbraidconcepttoolpreparing}. Delivering a magic state to where it is needed is therefore a \emph{co-optimization problem}: the compiler must decide where each operation's data patches and magic-state resource are located relative to each other, which compatible ports connect them, which route and timestep allow the operation to execute without conflicting with every other operation sharing the same layout. 

Crucially, these decisions cannot be made independently: two operations that need the same port, route, or magic-state terminal cannot both proceed, and the compiler has only two valid options: delay one of them, which lengthens execution, or keep a separate resource ready for each in advance, which enlarges the layout. Every operation in the program forces this same choice, which is why placement, routing, timing, and footprint have to be reasoned about together rather than one at a time.

\begin{figure}[t]
    \centering
    \includegraphics[width=0.8\linewidth]{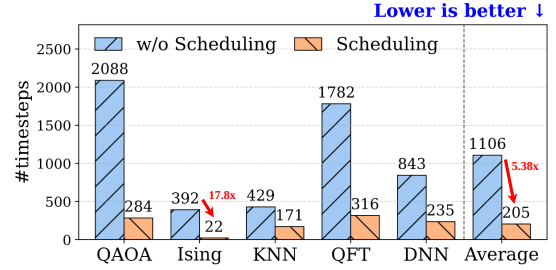}
    \caption{Number of timesteps with and without magic state scheduling. \textit{Scheduling lowers the execution time by an average 5.38$\times$ (up to 17.8$\times$).}}
    \label{fig:intro_plots}
\end{figure}

Across the whole program, these choices per operation accumulate into two costs. First, execution must be fast: every lattice-surgery operation occupies one or more \emph{logical timesteps}, each corresponding to several rounds of syndrome extraction that a real-time decoder must keep pace with to avoid stalling execution~\cite{das2022afs,decoding_latency_influence}, so the number of timesteps a program requires directly sets the wall-clock cost and the noise exposure of the computation~\cite{AharonovBenOr2008}, and finding a short schedule is itself hard~\cite{Herr_2017}. Fig.~\ref{fig:intro_plots} shows why this matters in practice: exploiting the parallelism available across operations, and taking the routing of operations into consideration, cuts the average number of timesteps across specific circuits by 5.38$\times$, and by up to 17.8$\times$. Second, that parallelism is not free: exposing it requires a layout with enough routing and magic-state capacity to realize it~\cite{kobori_lsqca_2025}, so a layout generous enough to make a short schedule reachable is, by construction, larger than what that schedule ends up using; execution time and layout decisions cannot be minimized independently.

Existing lattice-surgery compilers tend to pick one side of this trade-off~\cite{silva_multi-qubit_2024,hofmeyr_scheduling_2026}: a compiler that only tracks timesteps can cut execution time by letting the layout grow to whatever its routes need, while one that only tracks footprint can shrink the layout at the cost of execution time. Near-term fault-tolerant hardware can afford neither extreme: too few physical qubits to waste on an oversized layout, and too little runtime budget to waste on a schedule that sacrifices speed to save space. This gap leads us to the following research question:

\researchquestion{How can a compiler co-optimize the logical execution time and layout footprint of a lattice-surgery program, independent of the underlying magic-state generation protocol?}

Meeting this bar raises three concrete challenges, which we return to in \S\ref{sec:motivation}. First, \textit{execution time}: two operations with no dependency between them can still contend for the same data-patch ports, routing patches, or magic-state terminal once placed on the layout. Moreover, a placement that ignores frequent qubit interactions increases routing distance and, consequently, contention precisely where the circuit offers the most parallelism. Second, \textit{resource efficiency}: the layout must expose enough routing patches and magic-state terminals for the scheduler to find a conflict-free route for every candidate operation, so the initial layout is, by construction, a superset of what any single schedule ends up using; reporting it as the program's footprint overstates the resources execution actually required. Third, \textit{generality}: distillation supplies magic states from a fixed set of dedicated factories, while cultivation exposes smaller, distributed preparation sites with variable latency, so a scheduler whose resource model hard-codes one availability pattern cannot be reused to evaluate the other without rewriting its constraints.

We present Harvest, a resource-aware compiler for lattice-surgery that schedules magic-state consumption jointly with routing, placement, and layout footprint, independent of the magic state protocol. Its design follows three goals addressing these challenges: Harvest is \emph{performant}, packing operations that are dependency-ready, magic-state-available, and route-compatible into as few timesteps as possible; \emph{resource-efficient}, removing layout resources a schedule never touches rather than reporting the conservative footprint it started with; and \emph{general}, treating magic-state generation as a configurable resource model rather than a fixed protocol assumption.

Harvest realizes these goals as one compiler pipeline built around \emph{Harvest IR (H-IR)}, a patch-level intermediate representation that records operation dependencies, layout resources, and the final instruction schedule. It lowers input circuits into a dependency-aware Pauli-product representation, uses a \emph{circuit-aware layout constructor} to place logical qubits and allocate routing and magic-state patches, a \emph{routing-aware scheduler} to assign operations to timesteps under its magic-state resource model, and a \emph{post-scheduling layout pruner} to remove unused resources from the final layout.

We evaluate Harvest on QAOA, QFT, Feynman, Square-Heisenberg, and circuits from the QASMBench Small, Medium, and Large categories ~\cite{cross2017openquantumassemblylanguage,crossOpenQASM3Broader2022,li2022qasmbenchlowlevelqasmbenchmark}. Resource-aware scheduling reduces average schedule length from 612 to 127 logical timesteps, a 79.2\% reduction and a \(4.83\times\) average speedup over sequential execution, with individual benchmarks reaching up to \(17.8\times\). Circuit-aware placement improves speedup by up to \(1.35\times\) when routing locality limits the critical path, and post-scheduling pruning removes up to 72.0\% of unused magic-state patches and up to 33.9\% of unused routing patches for the most over-provisioned benchmark family. 


\myparagraph{Contributions} This paper makes the following contributions:

\begin{itemize}
    \item We formulate lattice-surgery code generation as a resource-constrained scheduling problem over data patches, routing patches, and magic-state    resources.

    \item We introduce Harvest, a modular compiler pipeline that combines circuit-aware layout construction, routing-aware scheduling, configurable magic-state modeling, and post-scheduling layout pruning.

    \item We define H-IR, a patch-level IR that records operation dependencies, layout resources, selected routes, magic-state use, and the final instruction schedule.

\end{itemize}

\begin{figure}[t]
    \centering
    \includegraphics[width=\linewidth]{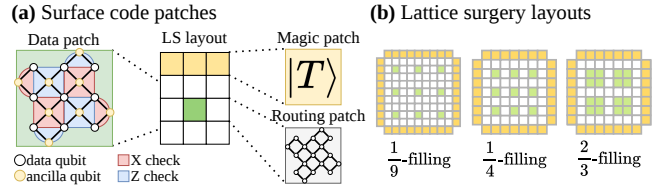}
    \caption{Patch-level execution model (\S\ref{sec:patch-level-fault-tolerant-execution}). \textit{\textbf{(a)} A physical surface-code patch is abstracted into a data, routing, or magic-state patch on a lattice-surgery layout. \textbf{(b)} Layout templates with different data-patch fillings.}}
    \label{fig:surface_code_patch}
\end{figure}

\section{Background and Motivation}
\label{sec:background}
This section defines the patch-level execution model used by Harvest (\S\ref{sec:patch-level-fault-tolerant-execution}), lattice-surgery operations (\S\ref{sec:lattice-surgery-operations}), magic-state resources (\S\ref{sec:magic-states-as-compiler-resources}), and finally the problem statement that this work addresses (\S\ref{sec:problem-statement}).
\subsection{Patch-Level Fault-Tolerant Execution}
\label{sec:patch-level-fault-tolerant-execution}

Surface codes protect a logical qubit by encoding it across a two-dimensional array of physical qubits and repeatedly measuring stabilizers to detect and correct errors~\cite{DennisEtAl2002,Fowler_2012}, and recent devices demonstrate this protection below the error-correction threshold~\cite{Acharya2025}. 
The resulting encoded unit, a \textit{patch}, is a contiguous block of physical qubits that realizes one logical degree of freedom. On a lattice-surgery layout, a patch takes one of three roles: a \textit{data patch} stores a logical qubit, a \textit{magic-state patch} holds a non-Clifford resource state, and a \textit{routing patch} provides idle space that connects other patches. Fig.~\ref{fig:surface_code_patch} shows this abstraction used in the paper. Patches are not static, however: they move and merge to carry out lattice-surgery operations (\S\ref{sec:lattice-surgery-operations}), and because magic states are costly to produce, their area must be used efficiently~\cite{litinski_game_2019,Gidney_2019}.



\subsection{Lattice-Surgery Operations}
\label{sec:lattice-surgery-operations}

Because each patch independently encodes one logical qubit, a multi-qubit operation requires interacting separately encoded patches without breaking their error protection. Lattice surgery implements logical operations by temporarily merging and splitting neighboring surface-code patches~\cite{HorsmanEtAl2012,bombin2006topological,litinski_game_2019}. From a compiler's perspective, each operation is a Pauli-product measurement: which logical qubits participate and which Pauli operator acts on each. For example, \(X \otimes I \otimes Z\) acts on the first and third qubits only. 

\myparagraph{Logical timesteps} A logical timestep is the scheduling unit for lattice-surgery operations: all involved patches merge, hold, and split within one timestep, corresponding physically to several rounds of syndrome extraction~\cite{HorsmanEtAl2012}.

\myparagraph{Requirements} Executing a Pauli product requires more than its logical operands: each data patch exposes \textit{ports} on its boundary, and an \(X\) term must connect to an \(X\) port, a \(Z\) term to a \(Z\) port. Non-adjacent patches connect through routing patches, occupied exclusively for the timestep. Together, these turn a Pauli-product operation into a spatial resource request: data patches, compatible ports, a route, and, for non-Clifford terms, a magic-state terminal. Two independent operations can therefore still conflict if they compete for the same patch-level resource.

\subsection{Magic States as Compiler Resources}
\label{sec:magic-states-as-compiler-resources}

Lattice-surgery operations are Clifford operations, which can be simulated efficiently classically and therefore cannot alone support universal computation~\cite{AaronsonGottesman2004}; universal FTQC requires implementing non-Clifford operations, such as \(T\) gates, without leaving the protected code space.

\myparagraph{Magic states} Non-Clifford operations consume magic states: specially prepared resource states that, once available on the layout, let a \(T\) operation be applied through an ordinary Clifford lattice-surgery operation rather than a direct, non-fault-tolerant gate~\cite{Bravyi_2005,Bravyi_2012}. Magic states are produced separately through  distillation or cultivation, which trade off fidelity, footprint, and preparation latency differently~\cite{gidney2024magicstatecultivationgrowing,hofmeyr_scheduling_2026}.

\myparagraph{Scheduling interface} The interface exposed to scheduling is crucial: a magic state is a consumable resource with a location and an availability time. Distillation exposes a predictable supply from dedicated factories, while cultivation exposes smaller, distributed resources with variable latency, differences captured as configurable availability models.

\myparagraph{Coupled constraints} A non-Clifford operation can execute only when a compatible magic state is available and reachable without conflicting with other operations in the same timestep, so availability cannot be reasoned about separately from routing and scheduling. These constraints jointly define the scheduling problem this paper addresses.
\begin{figure}
    \centering
    \includegraphics[width=0.8\linewidth]{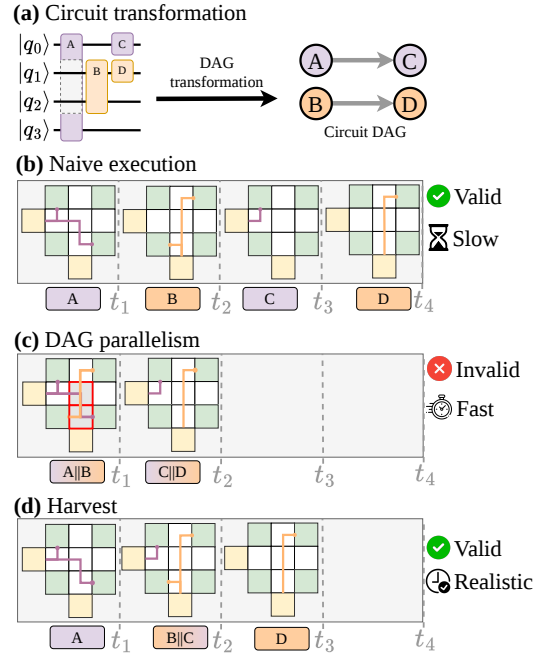}
    \caption{Resource-aware compilation motivation (\S\ref{sec:problem-statement}). \textit{\textbf{(a)} Circuit-to-DAG transformation. \textbf{(b)} Naive serial execution. \textbf{(c)} DAG-only scheduling. \textbf{(d)} Resource-aware scheduling.}}
    \label{fig:problem_figure}
\end{figure}

\begin{figure*}
    \centering
    \includegraphics[width=0.9\linewidth]{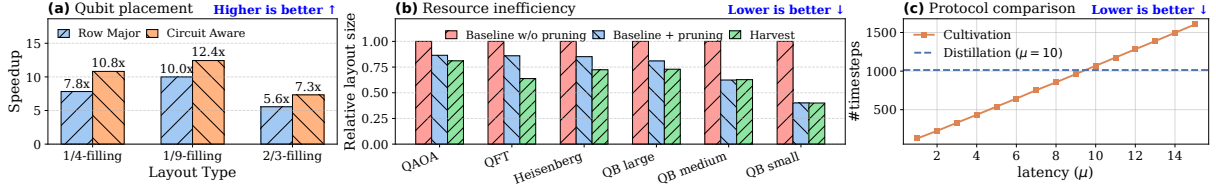}
    \caption{Three effects motivating resource-aware lattice-surgery compilation (\S\ref{sec:motivation}).
\textit{\textbf{(a)} Circuit-aware vs.\ row-major placement. \textbf{(b)} Patches removed by post-scheduling pruning. \textbf{(c)} Distillation vs.\ cultivation latency/throughput trade-offs.}}
    \label{fig:resource_aware_case}
\end{figure*}

\subsection{Motivation and Problem Statement}
\label{sec:problem-statement}
\myparagraph{Motivation} Circuit-level dependencies alone do not yield an efficient, valid schedule. Fig.~\ref{fig:problem_figure} shows this on a four-operation example: the circuit becomes a dependency DAG (a); a fully serial schedule is valid but slow (b); scheduling by DAG parallelism alone is fast but invalid, since independent operations can still conflict over a shared patch-level resource (c); only a schedule respecting both dependencies and resource constraints is valid and fast (d).

\myparagraph{Compiler input} The input is a DAG \(G_O=(V,E)\), a layout graph \(G_L=(V_L,E_L)\), and a magic-state resource model \(M\): nodes in \(G_O\) are Pauli-product operations with dependency edges; nodes in \(G_L\) are resources such as data-patch ports, routing patches, and magic-state terminals; and \(M\) describes when and where non-Clifford resources can be consumed.

\myparagraph{Compiler output} The output is a patch-level schedule, of the form shown in Fig.~\ref{fig:problem_figure}(d): for each operation \(v\), it assigns a logical timestep \(t(v)\), compatible data-patch ports, an optional magic-state terminal, and a route through the layout graph. A schedule is \textit{valid} only if it satisfies three conditions: every dependency edge in \(G_O\) is respected; every pair of operations sharing a timestep reserves disjoint resources from \(G_L\); and every magic state an operation consumes is available, at that location and timestep, under the resource model \(M\).

\statementLeftAccent{Problem Statement}{orange}{Given a quantum circuit, a lattice-surgery layout, and a magic-state resource model, how can a compiler generate the shortest \emph{valid} instruction schedule?}




\section{A Case for Resource-Aware Compilation}
\label{sec:motivation}


Finding a solution to the problem formulated in Section~\ref{sec:problem-statement}, however, depends on decisions the scheduler alone does not make: how qubits are placed on the layout before scheduling, how much routing and magic-state capacity the layout should provide, and how the choice of magic-state protocol is exposed to the scheduler. These decisions give rise to the three challenges addressed below.

\myparagraph{Challenge 1: Poor qubit placement}
The schedule depends on where logical qubits are placed. A row-major placement is simple and deterministic, but it ignores which qubits interact frequently in the Pauli-product graph. This can place strongly interacting qubits far apart, which increases route lengths and creates additional contention for routing patches. At the same time, placement cannot fully determine the final schedule: the scheduler still has to choose ports, routes, and compatible operations for each timestep. Harvest uses circuit information to improve the initial layout (Fig.~\ref{fig:resource_aware_case}(a)), while leaving the exact routing decisions to the scheduler.

\statementLeftAccent{Key idea \#1}{blue}{\textbf{Circuit-aware placement.} Use the Pauli-product operation graph to guide qubit placement before scheduling, so that frequently interacting qubits are placed closer together while leaving the scheduler free to resolve concrete routing conflicts.}

\myparagraph{Challenge 2: Resource inefficiency}
The layout used during scheduling should provide enough routing and magic-state resources to make efficient schedules possible. If the layout is too small, valid routes may be unavailable even when the operation DAG has parallelism. However, the final schedule usually uses only part of the initially allocated auxiliary space (Fig.~\ref{fig:resource_aware_case}(b)), so reporting the initial layout as the final footprint would therefore overestimate the resources used by the compiled program. Harvest separates these two roles: the initial layout serves as the scheduling search space, and the final layout is derived from the resources actually used in the routed schedule.

\statementLeftAccent{Key idea \#2}{blue}{\textbf{Layout pruning.} Compile with a flexible layout, then prune only after scheduling by keeping the patches that are actually referenced by routed instructions in the final instruction schedule.}
\begin{figure*}
    \centering
    \includegraphics[width=\linewidth]{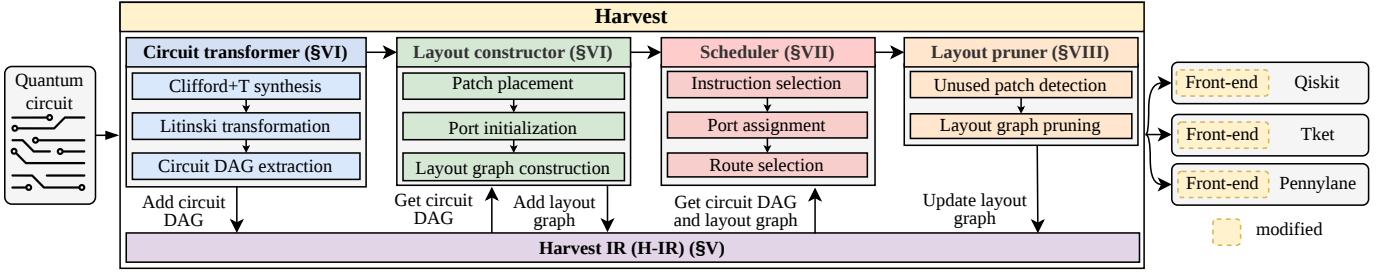}
    \caption{Harvest architecture (\S\ref{sec:architecture}). \textit{Harvest sits between circuit-level frontends and lower-level lattice-surgery backends. It lowers input circuits to Pauli-product operations, constructs a logical lattice-surgery layout, schedules operations under routing and magic-state constraints, prunes unused auxiliary resources, and emits H-IR as a routed patch-level program.}}
    \label{fig:Harvest_postion}
\end{figure*}

\myparagraph{Challenge 3: Lack of generality}
Magic-state generation affects scheduling through both time and location. Distillation and cultivation provide magic states with different latency, throughput, and footprint trade-offs (Fig.~\ref{fig:resource_aware_case}(c)). A scheduler that hard-codes one protocol cannot easily evaluate another, while a model that only counts available magic states loses the information needed for routing. Harvest treats magic-state generation as a resource model with compiler-visible availability: where a magic state can be consumed, when it becomes available, and whether it has already been used in the current schedule.

\statementLeftAccent{Key idea \#3}{blue}{\textbf{Protocol-agnostic magic states.} Represent magic-state generation as a configurable resource model, so the same scheduler can reason about different protocols through their compiler-visible interface: when and where magic states become available.}

Together, these key ideas motivate Harvest's design choices: keep dependencies, placement, routing, magic-state availability, and final layout usage explicit throughout compilation.

\section{Overview}
\label{sec:overview}

Harvest is a compiler for scheduling lattice-surgery instructions under routing and magic-state constraints. Its input is a quantum circuit, and its output is a scheduled patch-level program expressed over the abstraction from \S\ref{sec:background}: data patches, routing patches, magic-state resources, and logical timesteps.

\subsection{Harvest Architecture}
\label{sec:architecture}

Fig.~\ref{fig:Harvest_postion} shows Harvest's position in the compilation stack: four components built around a shared intermediate representation, H-IR.

\myparagraph{Input and output} Harvest's own decisions should stay decoupled from hardware-specific lowering: it takes a quantum circuit as input and produces H-IR, a routed patch-level IR, as output.

\myparagraph{Harvest IR (H-IR)} Because all four components must reason over the same dependencies, layout, and schedule, H-IR serves as their shared representation.

\myparagraph{Circuit transformer} Since the rest of the pipeline needs the circuit's dependencies and lattice-surgery structure exposed up front, the circuit transformer produces three representations of the input program: a \textit{Clifford+T} decomposition, a \textit{Pauli-product} rewrite, and the \textit{operation DAG} \(G_O\) capturing dependencies between operations.

\myparagraph{Layout constructor} Before any operation is placed, the scheduler needs a spatial search space to route within -- the layout constructor builds it in three steps: \textit{layout construction} places data patches, \textit{port initialization} exposes each patch's data, routing, and magic-state ports, and \textit{graph construction} connects them into the layout graph \(G_L\).

\myparagraph{Scheduler} An operation should advance only once it is simultaneously dependency-ready, resource-compatible, and routable; the scheduler enforces this through three modules: \textit{instruction selection} picks dependency-ready operations, \textit{port assignment} assigns compatible data and magic-state ports, and \textit{route selection} finds a conflict-free route through \(G_L\).

\myparagraph{Layout pruner} A useful layout reflects actual use, not the conservative initial layout, which is why the layout pruner finalizes the layout in two steps: \textit{unused-patch detection} identifies patches and ports that no scheduled instruction used, and \textit{graph pruning} removes them from \(G_L\).

\subsection{Harvest Workflow}
\label{sec:workflow}

These four components transform a quantum circuit into a routed lattice surgery instruction schedule, as shown in Fig.~\ref{fig:Harvest_postion}.

\myparagraph{Transformation and analysis} A quantum circuit enters the circuit transformer, which produces the operation DAG together with the qubit-interaction, operation-size, and magic-state-demand information the layout constructor needs.

\myparagraph{Layout construction} The layout constructor consumes this information to place logical qubits on data patches and allocate auxiliary routing and magic-state patches; placement shapes route lengths and therefore routing contention during scheduling.

\myparagraph{Scheduling} The scheduler repeatedly selects ready operations and places compatible ones in the same logical timestep, reserving data patches, magic-state resources, and a route for each, combining list scheduling with spatial routing and magic-state allocation.

\myparagraph{Pruning} Once scheduling finishes, the pruner removes routing and magic-state patches the final schedule never used, so the layout reported in H-IR reflects the resources actually needed rather than the initial search space.

This modular structure lets placement, scheduling, magic-state modeling, and pruning be evaluated and replaced independently, since the best design choice depends on the interaction between program structure, routing pressure, and magic-state availability rather than on any single component in isolation.

\subsection{Harvest Resource Model}
\label{sec:resource-model}

The Harvest components coordinate through a shared, explicit view of the machine's resources: the scheduler can only pack operations, and the pruner can only remove patches, over resources the compiler represents. We define this resource model, whose spatial side is the layout graph and whose temporal side is magic-state availability; H-IR records both (\S\ref{sec:h-ir}) and the scheduler treats them as constraints (\S\ref{sec:scheduling}).

The layout graph $G_L=(V_L,E_L)$ represents the lattice-surgery layout. Its nodes include data-patch ports, routing patches, and magic-state terminals, while its edges describe connections between these resources. Data patches are exposed through ports because different boundaries support different logical Pauli operators. Routing patches provide connectivity, and magic-state patches are regions from which non-Clifford operations consume a magic state. Magic-state availability is modeled separately. We represent the scheduling interface as
\begin{equation*}
    M \subseteq V_M \times \mathbb{N}_{\geq 0},
    \label{eq:hir-magic-model}
\end{equation*}
where $V_M\subseteq V_L$ is the set of magic-state terminals. This interface allows different generation protocols to expose different availability patterns with the same scheduler interface.

\section{Harvest Intermediate Representation (H-IR)}
\label{sec:h-ir}

 \begin{figure*}
    \centering
    \includegraphics[width=\linewidth]{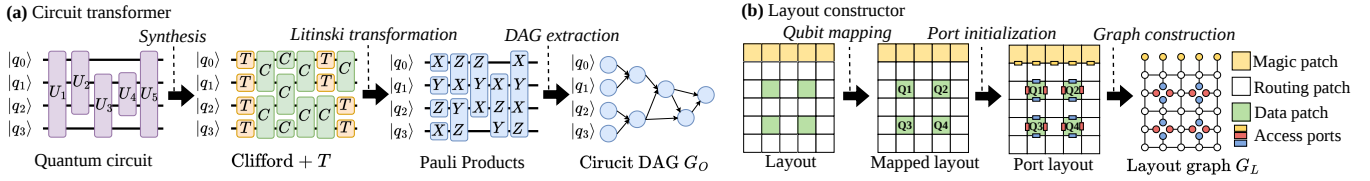}
    \vspace{-12pt}
    \caption{Transformation workflows (§\ref{sec:CT_GC}). \textit{(a) The circuit transformer turns a quantum circuit into the circuit operation DAG $G_O$. (b) The layout constructor builds the layout graph $G_L$.}}
     \vspace{-6pt}
    \label{fig:cirucit_transformer_gc}
\end{figure*}

Harvest IR (H-IR) is the interface between circuit-level semantics and resource-aware lattice-surgery code generation. At the circuit level, an operation is defined by its operands and dependencies. Lattice-surgery execution additionally requires compatible patch boundaries, routing resources, and, for non-Clifford operations, a magic state available at the required time. H-IR keeps these decisions in a common logical patch-level representation, tying together the operation DAG $G_O$, the layout graph $G_L$, the magic-state model $M$, and the resulting instruction schedule $S$, following the same multi-level design philosophy as general-purpose compiler infrastructures~\cite{lattnerMLIRCompilerInfrastructure2020} but specialized to lattice-surgery resources. H-IR operates above physical surface-code execution and therefore abstracts physical qubits, syndrome extraction, decoding~\cite{das2022afs,pymatching,sliding_window}, and hardware control.

\myparagraph{Operation semantics}
The operation DAG $G_O$ (\S\ref{sec:problem-statement}) represents the transformed program as Pauli-product operations connected by dependency edges inherited from circuit transformation. Each node additionally records which logical qubits participate and which Pauli operator acts on each, since this is what port assignment consults during scheduling. Operations without a dependency path are candidates for parallel execution, but their simultaneous execution still depends on the available layout and magic-state resources: the DAG therefore separates semantic dependencies from the resource conflicts introduced during code generation.

\myparagraph{Scheduled instructions}
The instruction schedule $S$ records how operations are realized on the layout. We represent a scheduled instruction as $ s = \langle v, t, P_s, m_s, R_s \rangle$
where $v$ is the corresponding Pauli-product operation, $t$ its logical timestep, $P_s$ the selected data-patch ports, $m_s$ an optional magic-state terminal, and $R_s$ the reserved routing resources. The operation refers back to $G_O$, while the ports, terminal, and route refer to resources in $G_L$: H-IR therefore captures both what operation executes and the resources used to realize it. Table~\ref{tab:hir-schedule-entry} shows a simplified schedule entry in H-IR.

\begin{table}[t]
\centering
\caption{Simplified schedule entry example in H-IR (\S\ref{sec:h-ir}).}
\label{tab:hir-schedule-entry}
\small
\begin{tabular}{@{}ll@{}}
\toprule
\textbf{Field} & \textbf{Value} \\
\midrule
Timestep          & $0$ \\
Qubit indices     & $[0]$ \\
Magic terminal    & \texttt{P:mL2:M\_E:M} \\
Qubit ports       & \texttt{P:q\_0:W:Z} \\
Routing cells     & $(1,2)$ \\
Port nodes        & \texttt{P:mL2:M\_E:M}, \texttt{P:q\_0:W:Z} \\
\bottomrule
\end{tabular}
\end{table}

\myparagraph{H-IR invariants}
Beyond the dependency, resource-disjointness, and magic-state-availability conditions already required of a valid schedule (\S\ref{sec:problem-statement}), a completed H-IR instance must also route every non-Clifford instruction's selected terminals through a connected route in $G_L$. Keeping $G_O$, $G_L$, and $M$ explicit and separate lets H-IR also preserve, rather than discard, the information needed to tell apart dependency-, routing-, and magic-state-limited execution (\S\ref{sec:scheduling}). These invariants form the contract between the scheduler and the downstream stages that consume H-IR.

\myparagraph{Compiler interface}
H-IR is assembled incrementally across the pipeline introduced in \S\ref{sec:overview}: circuit transformation and layout construction populate $G_O$ and $G_L$, scheduling produces $S$ against $M$, and pruning refines $G_L$ to the footprint the schedule actually realized. Keeping all of these explicit in one representation, rather than passing ad hoc state between stages, is what lets placement, routing, scheduling, and magic-state availability be reasoned about jointly rather than in isolation.

\section{Transformer and Layout Constructor}
\label{sec:CT_GC}

The scheduler consumes two inputs that the input circuit does not directly provide: the operation DAG $G_O$, which defines \emph{what} may execute, and the layout graph $G_L$, which defines \emph{where}. The circuit transformer produces $G_O$, and the layout constructor produces $G_L$ in two steps: it places logical qubits using the circuit's interaction structure, then builds the spatial substrate around that placement.

\myparagraph{Circuit transformation}
Fig.~\ref{fig:cirucit_transformer_gc}(a) shows how Harvest converts a
quantum circuit into the DAG representation used by the scheduler. Harvest first lowers the circuit to the Clifford+$T$ gate set, then rewrites it into Pauli-product operations using the Litinski transformation~\cite{litinski_game_2019}, and finally constructs the operation DAG $G_O$, whose nodes represent Pauli-product operations and whose edges encode their execution dependencies. The same Pauli-product representation also yields the qubit-interaction information that guides placement.

\begin{figure}
    \centering
    \includegraphics[width=0.7\linewidth]{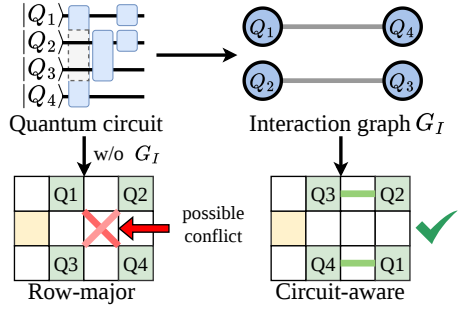}
    \caption{Qubit placement (§\ref{sec:CT_GC}). \textit{Row-major placement follows layout order, while circuit-aware placement uses the qubit interaction graph to improve locality between frequently interacting qubits.}}
    \label{fig:circuit-aware-mapping}
\end{figure}

\myparagraph{Circuit-aware placement}
\label{sec:placement}
The position of a data patch affects the routes available to every operation on its logical qubit. A row-major placement assigns qubits to data patches by index and provides a simple deterministic baseline, but it ignores the interaction structure of the program. Harvest instead biases the initial placement toward frequently interacting qubits to reduce expected routing constraints, as illustrated in Fig.~\ref{fig:circuit-aware-mapping}.

We represent circuit structure as a weighted interaction graph $G_I$ over the logical qubits $Q$. For each pair $(q_i,q_j)$, the weight $w_{ij}$ counts how often both qubits participate in the same Pauli-product operation. Given the available data-patch locations $P$ and the layout distance $d(p_a,p_b)$ between two locations, Harvest guides the placement $\pi:Q\rightarrow P$ with the objective

\begin{equation*}
    C(\pi)=\sum_{i,j} w_{ij}\,d\!\left(\pi(q_i),\pi(q_j)\right).
    \label{eq:placement-objective}
\end{equation*}

Lower values place strongly interacting qubits closer together. Harvest uses this objective as a heuristic guide rather than solving the assignment problem optimally: the placement pass prioritizes qubits with strong interaction profiles, assigns them to favorable data-patch locations, and produces the qubit mapping that layout construction consumes.

The placement objective captures locality, but not the complete scheduling problem. Shorter interaction distances reduce route lengths and contention. The final schedule still depends on operation dependencies, magic-state availability, and conflicts between simultaneously routed operations. Circuit-aware placement gives the scheduler a better starting point without attempting to predict the complete execution.

\myparagraph{Layout-graph construction}
Fig.~\ref{fig:cirucit_transformer_gc}(b) illustrates the final step. Once the qubit mapping is fixed, Harvest instantiates the selected layout template and assigns each logical qubit to its data patch. It then adds routing and magic-state patches according to the selected configuration, exposes $X$ and $Z$ patch boundaries as ports, and converts the resulting grid into the layout graph $G_L$, which encodes the layout resources and their connectivity and, together with $G_O$, the input to the scheduler (\S\ref{sec:scheduling}).

\section{Resource-Aware Scheduler}
\label{sec:scheduling}

The scheduler consumes the operation DAG $G_O$ and layout graph $G_L$ (\S\ref{sec:CT_GC}), together with the magic-state model $M$ (\S\ref{sec:resource-model}), and produces the instruction schedule $S$. Scheduling combines three constraints: operation dependencies, magic-state availability, and routing resources. An operation may be ready in $G_O$ yet unable to execute if its magic state or routing resources are unavailable. We therefore construct each logical timestep from the operations that are both logically ready and spatially realizable on the current layout.

\myparagraph{Constructing a timestep}
We treat each logical timestep as a temporary resource-allocation problem. Let $P$ be the set of already scheduled operations. At timestep $t$, we first compute the ready set

\begin{equation*}
    E_t =
    \left\{
        v \in V \setminus P
        \;\middle|\;
        \operatorname{pred}(v) \subseteq P
    \right\}.
    \label{eq:ready-set}
\end{equation*}


We then check whether the required magic-state resources are available. Non-Clifford operations require an available compatible magic-state patch. The parameter $\mu$ denotes the nominal preparation latency. Distillation uses a deterministic latency of exactly $\mu$ timesteps, whereas cultivation uses a variable latency bounded by $\mu_{\min}$ and $\mu_{\max}$ around $\mu$. The scheduler consumes a selected state only after the operation has been routed successfully and committed to the schedule. Operations that fail to route remain eligible for later timesteps.

\myparagraph{Materializing spatial demand}
For each candidate operation, we select the layout terminals required to realize its Pauli product. An $X$ or $Z$ term requires a compatible boundary of the corresponding data patch, while a $Y$ term requires access to both boundary types. Non-Clifford operations additionally include a selected magic-state terminal. Candidates that are independent in the operation DAG may still compete for the same magic-state terminal, or routing region. The final set of operations for a timestep depends on whether their routing requests can be realized concurrently.

\myparagraph{Routing candidate operations}
An operation may connect more than two terminals. We construct an approximate Steiner tree over the layout graph for each candidate and use the resulting tree as its tentative route. We route candidates independently to expose available parallelism. The resulting routes may overlap. Let $\rho_t(r)$ denote the number of tentative routes using routing resource $r$ and let $c(r)$ denote its capacity. We quantify the resulting routing contention as

\begin{equation*}
    \Phi_t =
    \sum_{r \in \mathcal{R}}
    \max\left(0, \rho_t(r) - c(r)\right).
    \label{eq:routing-conflict}
\end{equation*}

In our patch-level model, routing resources have unit capacity. A timestep is
therefore conflict-free when $\Phi_t = 0$.

\myparagraph{Negotiating congestion}
When tentative routes overlap, we iteratively reroute the operations involved in the
conflict. The routing cost combines current congestion with a history term,

\begin{equation*}
    w_t(r)
    =
    \alpha \frac{\rho_t(r)}{c(r)}
    +
    \beta h_t(r),
    \label{eq:negotiated-congestion}
\end{equation*}

where $h_t(r)$ records how often routing resource $r$ has remained congested across previous iterations. Current congestion steers routes away from occupied resources, while the history term discourages repeated use of persistent bottlenecks. We continue rerouting until all conflicts are resolved or the rerouting budget is exhausted. If conflicts remain, we defer the operations contributing most to the remaining congestion. Deferred operations remain unscheduled and are reconsidered in a later timestep. Algorithm~\ref{alg:harvest-pack} summarizes the construction of one logical timestep.

\begin{algorithm}[t]
\caption{Conflict-aware timestep construction}
\label{alg:harvest-pack}
\small
\begin{algorithmic}[1]
\Require Candidate operations $E_t$, layout graph $G_L$
\Ensure Conflict-free scheduled subset $A_t$

\Statex \Comment{Turn each Pauli product into a concrete resource request}
\State select compatible terminals for all $v \in E_t$
\Statex \Comment{Route candidates independently to expose parallelism}
\State route each candidate using an approximate Steiner tree
\Statex \Comment{Quantify contention across candidate routes}
\State compute routing usage $\rho$ and conflict score $\Phi$

\Statex \Comment{Iteratively negotiate away routing conflicts}
\While{$\Phi > 0$ and rerouting budget remains}
    \State identify over-capacity routing resources
    \State identify routes using these resources
    \ForAll{conflicting routes $R_i$}
        \Statex \Comment{Steer around congested and historically busy resources}
        \State reroute $R_i$ using
        $w(r)=\alpha\rho(r)/c(r)+\beta h(r)$
    \EndFor
    \State update congestion history $h$, routing usage $\rho$, and $\Phi$
\EndWhile

\Statex \Comment{Give up and push the worst offender to a later timestep}
\While{$\Phi > 0$}
    \State defer the operation contributing most to congestion
    \State update $\rho$ and $\Phi$
\EndWhile

\State \Return operations with remaining routes
\end{algorithmic}
\end{algorithm}

\begin{figure*}
    \centering
    \includegraphics[width=\linewidth]{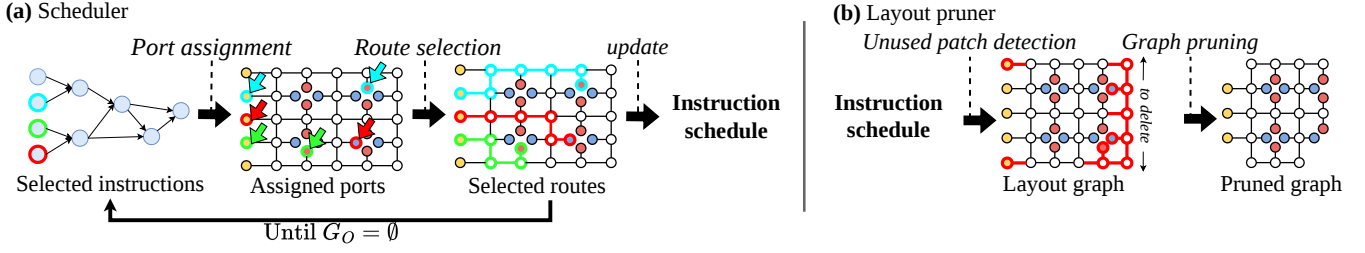}
     \vspace{-20pt}
    \caption{Scheduling (§\ref{sec:scheduling}) and layout pruning (§\ref{sec:pruning}) workflow. \textit{(a) The scheduler selects ready instructions, assigns data and magic-state ports, and routes each operation through the layout. (b) The pruner marks layout resources referenced by the routed schedule and removes the rest from the layout graph.}}
     \vspace{-5pt}
    \label{fig:scheduler_pruner}
\end{figure*}

\myparagraph{Committing the timestep}
Once routing produces a conflict-free subset $A_t$, we assign these operations to timestep $t$. We record their selected data ports, routes, and magic-state terminals, consume the corresponding magic states, and mark the operations as scheduled. Routing resources are released before constructing the next timestep and the scheduler repeats this process until all operations have been scheduled. If no ready operation can execute at the current timestep, we advance the resource model and continue with the next timestep.

\myparagraph{Exposing scheduling bottlenecks}
The resulting schedule reflects three possible limiting factors. A workload is \emph{dependency-limited} when too few operations reach the frontier, \emph{magic-state-limited} when ready non-Clifford operations lack an available resource state, and \emph{routing-limited} when otherwise executable operations cannot be placed on disjoint routing resources. Keeping these constraints visible matters because they interact. Additional magic-state capacity provides little benefit when routing is the dominant bottleneck, while reducing route lengths cannot expose parallelism hidden by the dependency graph. Harvest stores the final routed schedule in H-IR, recording for each operation the
execution timestep, selected data ports, optional magic-state terminal, and reserved routing resources. 

\section{Layout Pruner}
\label{sec:placement-pruning}
\label{sec:pruning}
Every scheduled instruction records the ports, route, and magic-state terminal it reserved (\S\ref{sec:scheduling}), so the completed schedule reveals which of the layout's routing resources the program actually used. Harvest exploits this to optimize the layout in two ways: circuit-aware placement (\S\ref{sec:CT_GC}) shapes the routing problem before scheduling, while post-scheduling pruning reclaims the unused patches afterward.


\myparagraph{Bounding the initial layout}
The initial layout serves as a search space for scheduling. At construction time, the compiler does not yet know which routing cells or magic-state resources the scheduler will select, so the layout may contain routing resources that no instruction eventually uses. Retaining all of these resources in the final H-IR would report the compiler's initial search space rather than the footprint realized by the generated schedule.

\myparagraph{Computing the pruned layout}
Harvest prunes the layout after routing and scheduling are complete. Let $U(S)=\bigcup_{s\in S}\operatorname{res}(s)$ denote the resources referenced by at least one scheduled instruction, where $\operatorname{res}(s)$ collects the data ports $P_s$, magic-state terminal $m_s$, and routing resources $R_s$ that instruction $s$ records in H-IR (\S\ref{sec:h-ir}). We retain all data resources $V_D\subseteq V_L$ together with the set:

\begin{equation*}
    V_L' = V_D \cup U(S),
    \qquad
    G_L' = G_L[V_L'].
    \label{eq:pruned-layout}
\end{equation*}

Unused routing resources, ports, and magic-state resources can therefore be removed where applicable, while data patches remain because they store the logical program state. Since every scheduled instruction records its selected ports, magic-state terminal, and route, all resources required by the generated schedule belong to $U(S)$. Removing auxiliary resources outside this set leaves every scheduled route unchanged.

\myparagraph{Pruning after scheduling}
Pruning deliberately follows scheduling. Removing auxiliary resources earlier would restrict the routing choices available to the scheduler and could prevent otherwise feasible parallel execution. The initial graph $G_L$ therefore represents the spatial search space available during compilation, while $G_L'$ represents the realized footprint of the generated schedule. This footprint is schedule-specific and is not claimed to be a globally minimum layout.
\begin{figure*}[t]
\centering
\includegraphics[width=\linewidth]{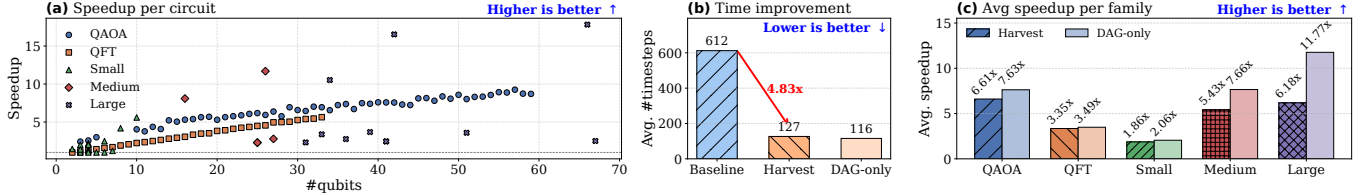}
\caption{End-to-End performance (\S\ref{sec:eval-scheduling}). \textit{\textbf{(a)} Speedup over sequential execution per benchmark circuit. \textbf{(b)} Average logical timesteps with and without scheduling. \textbf{(c)} Average speedup by circuit family and distance to ideal DAG parallelism.}}
\label{fig:rq_1}
\end{figure*}
Placement and pruning capture the spatial decisions surrounding Harvest's scheduler. Circuit-aware placement uses program structure to shape routing demand before scheduling, while pruning uses the completed routed schedule to remove unused auxiliary resources afterward. This separation allows the scheduler to explore a flexible layout without requiring the final H-IR to retain every resource initially provisioned. 

\section{Evaluation}
\label{sec:evaluation}

We evaluate Harvest at the logical patch level; the reported metrics describe the generated lattice-surgery schedules and layout resources. We follow the design goals from \S\ref{sec:overview}: \S\ref{sec:eval-scheduling} measures end-to-end performance, including how scheduling behaves under both distillation and cultivation; \S\ref{sec:eval-placement} isolates the contribution of circuit-aware placement; and \S\ref{sec:eval-pruning} measures the resource efficiency gained from layout pruning.

\begin{table}[t]
\centering
\caption{Benchmark dataset. \textit{Summary of the circuit families, number of circuits, logical-qubit range, and range of transformed Pauli product operations.}}
\label{tab:benchmark-dataset}
\footnotesize
\renewcommand{\arraystretch}{1.15}
\begin{tabularx}{\linewidth}{X r r r}
\toprule
\textbf{Family} &
\textbf{\#Circuits} &
\textbf{\#Qubits} &
\textbf{\#Pauli products} \\
\midrule
Feynman & 48 & 1--48 & 5--171{,}482 \\
QAOA & 53 & 3--59 & 51--1{,}227 \\
QFT & 32 & 2--33 & 15--1{,}782 \\
Square-Heisenberg & 11 & 4--64 & 316--8{,}800 \\
QASMBench large & 11 & 31--67 & 2--847 \\
QASMBench medium & 34 & 11--27 & 2--4{,}256 \\
QASMBench small & 51 & 2--10 & 1--3{,}117 \\
\bottomrule
\end{tabularx}
\end{table}

\subsection{Experimental Setup}
\label{sec:eval-setup}
We detail the experimental setup we use to evaluate Harvest. 

\myparagraph{Metrics}
We use logical timesteps, speedup over sequential execution, and patch reduction after pruning. Speedup is defined as \[S = \frac{\text{sequential timesteps}} {\text{scheduled timesteps}}.\]

\myparagraph{Benchmarks}
We evaluate Harvest on QASM benchmark circuits from \textit{QAOA}, \textit{QFT}, \textit{Clifford}, \textit{Feynman}, \textit{Square-Heisenberg}, and \textit{QASMBench Small}, \textit{Medium}, and \textit{Large} families, summarized in Table~\ref{tab:benchmark-dataset}. The benchmarks cover a range of logical qubit counts, Pauli-product graph sizes, and available circuit-level parallelism.
 
\myparagraph{Baselines}
We compare Harvest against two baselines. Sequential execution runs the Pauli-product operations in dependency order, while DAG-only execution schedules all independent operations in parallel without considering resource conflicts, providing an optimistic upper bound. All variants use the same transformed circuit, lattice-surgery layout, qubit placement, and magic-state configuration.

\myparagraph{Hardware setup}
We run all experiments on a Lenovo ThinkPad E16 Gen.~2 equipped with an AMD Ryzen 5 7000-series processor and 16\,GB of RAM on Ubuntu \texttt{24.04.4}.
Unless stated otherwise, we keep all parameters fixed when comparing compiler variants and change only the component under evaluation. 


\subsection{End-to-End Performance}
\label{sec:eval-scheduling}

We evaluate the end-to-end speedup from resource-aware scheduling, how that speedup varies across benchmark families, and how sensitive it is to magic-state availability.

\researchquestion[RQ1: Harvest speedup]{How much does resource-aware scheduling reduce logical execution time compared to sequential execution?}

\myparagraph{Methodology}
We compare Harvest against a sequential baseline. The baseline executes the transformed Pauli-product operations one after another in dependency order. Harvest uses the same transformed circuit, layout, placement, and magic-state configuration, but schedules independent operations in the same logical timestep when their data patches, routes, and magic-state resources do not conflict.

\myparagraph{Analysis}
Fig.~\ref{fig:rq_1}(a) reports the speedup over sequential execution for individual benchmark circuits, while Fig.~\ref{fig:rq_1}(b) compares average logical timesteps. Harvest reduces the average schedule length from 612 to 127 logical timesteps. This saves 485 timesteps on average, corresponding to a 79.2\% reduction and an average speedup of 4.83$\times$.

\statementLeftAccent{Takeaway \#1}{green}{Harvest cuts average schedule length by 79.2\% (612 $\rightarrow$ 127 timesteps), a 4.83$\times$ speedup over sequential execution.}

\researchquestion[RQ2: Workload sensitivity]{How does the benefit of resource-aware scheduling vary across circuit families?}

\myparagraph{Methodology}
We aggregate the scheduling results by circuit family. We focus on relative speedups and examine whether scheduling benefits are concentrated in specific circuit classes.

\begin{figure}
    \centering
    \includegraphics[width=\linewidth]{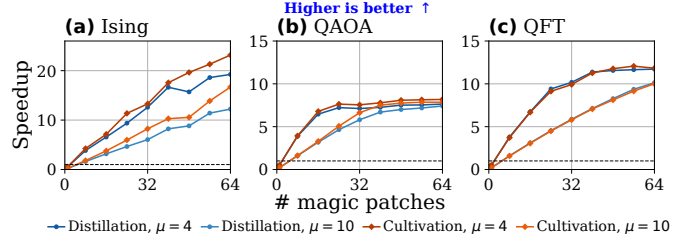}
    \caption{Magic-state availability (\S\ref{sec:eval-scheduling}). \textit{Average speedup for 25--31-qubit Ising, QAOA, and QFT circuits on a shared layout under distillation and cultivation ($\mu=4,10$).}}
    \label{fig:rq_3}
\end{figure}
\begin{figure*}[t]
    \centering
    \includegraphics[width=\linewidth]{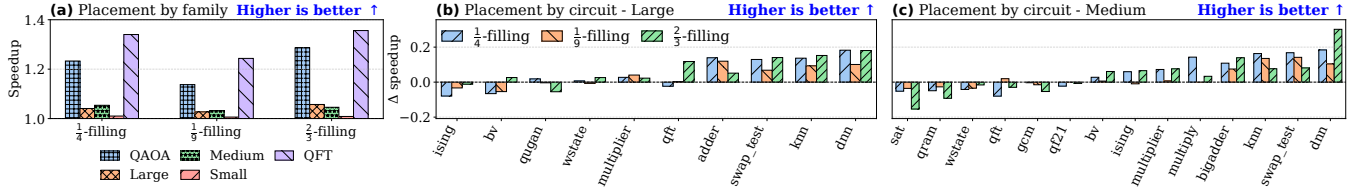}
    \caption{Placement effectiveness (\S\ref{sec:eval-placement}). \textit{\textbf{(a)} Placement speedup of circuit-aware placement over row-major placement for different layout templates under Harvest scheduling. Values above one indicate that circuit-aware placement reduces the number of logical timesteps. \textbf{(b)} and \textbf{(c)} per-circuit change in speedup for Large and Medium benchmarks, respectively.}}
\label{fig:rq_4_speedup_family}
\end{figure*}

\myparagraph{Analysis}
Fig.~\ref{fig:rq_1}(c) shows that Harvest captures most of the available parallel speedup. The gap is small for QFT (3.35$\times$ vs. 3.49$\times$), Small (1.86$\times$ vs. 2.06$\times$), and QAOA (6.61× vs. 7.63$\times$). The larger gaps for Medium and Large are mainly driven by a few outlier circuits, such as Ising, where routing and shared-resource conflicts limit realizable parallelism. For QAOA we get the largest speedup of 6.61 $\times$.


\statementLeftAccent{Takeaway \#2}{green}{Harvest benefits every benchmark family, achieving up to 6.61$\times$ average speedup. For QFT, Small, and QAOA, it realizes 96\%, 90\%, and 87\% of the DAG-only parallelism bound, respectively.}


\researchquestion[RQ3: Magic-state availability sensitivity]{How does schedule length scale with the number of available magic-state patches?}

\myparagraph{Methodology}
We average results for Ising, QAOA, and QFT circuits with 25--31 logical qubits, which share the same lattice-surgery layout. We vary the number of magic-state patches from 1 to 64 and evaluate distillation and cultivation with $\mu=4$ and $\mu=10$.

\myparagraph{Analysis}
Fig.~\ref{fig:rq_3} shows that additional magic-state patches improve speedup by enabling more non-Clifford operations to execute concurrently. QAOA and QFT begin to saturate once sufficient capacity is available, while Ising remains sensitive across the evaluated range. A larger $\mu$ delays saturation because each patch becomes available less frequently.

\statementLeftAccent{Takeaway \#3}{green}{With $\mu=4$, QAOA and QFT saturate at approximately 24 and 48 magic-state patches, reaching about $8\times$ and $12\times$ speedup. Ising does not saturate yet with 64 patches ($23\times$).}



\subsection{Placement Effectiveness}
\label{sec:eval-placement}

We evaluate how much circuit-aware qubit placement further reduces schedule length beyond scheduling alone, and how consistent that effect is across circuits.

\researchquestion[RQ4: Placement effectiveness]{How much does circuit-aware qubit placement reduce schedule length compared to row-major placement?}

\myparagraph{Methodology}
We compare circuit-aware placement against row-major placement. Row-major placement assigns logical qubits to data patches by index, without considering circuit structure (Fig.~\ref{fig:circuit-aware-mapping}). Circuit-aware placement instead uses the weighted qubit-interaction graph to place frequently interacting qubits closer together.
The scheduler, layout template, circuit, and magic-state model are kept fixed. Here the Speedup is calculated as \[S = \frac{T_{row}} {T_{circuit-aware}}.\]
\begin{figure*}[t]
    \centering
    \includegraphics[width=\linewidth]{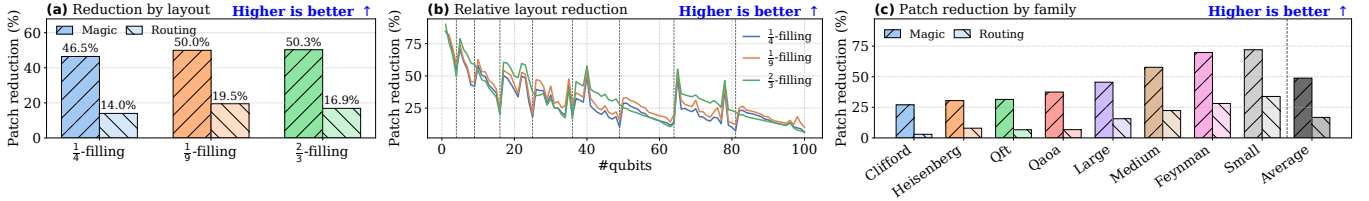}
    \caption{Resource efficiency (\S\ref{sec:eval-pruning}). \textit{\textbf{(a)} Patch reduction by layout template and resource type; \textbf{(b)} percentage of removed patches relative to the initial layout as circuit size changes; 
    and \textbf{(c)} patch reduction by circuit family.}}
    \label{fig:pacthes_removed_by_layout_line}
\end{figure*}

\myparagraph{Analysis}
Fig.~\ref{fig:rq_4_speedup_family}(a) shows placement speedup for different layout templates under Harvest scheduling. Values above one indicate that circuit-aware placement reduces the number of logical timesteps compared to row-major placement. The results show that placement can improve schedules when routing locality affects the critical path.

Fig.~\ref{fig:rq_4_speedup_family}(b) and (c) show the per-circuit placement effect for Large and Medium benchmarks. The effect is circuit-dependent. Some circuits improve, some remain almost unchanged, and some degrade slightly. This behavior is expected because placement changes routing distance and congestion, but it does not change the dependency graph or the number of available magic states.

\statementLeftAccent{Takeaway \#4}{green}{Circuit-aware placement improves speedup by up to 1.35$\times$, but only when routing congestion sets the critical path.}

\subsection{Resource Efficiency}
\label{sec:eval-pruning}

We evaluate how much auxiliary layout space post-scheduling pruning reclaims from the initial, conservative layout, and how that reclaimed fraction varies with circuit size, layout template, and benchmark family.

\researchquestion[RQ5: Layout footprint]{How much auxiliary layout space remains unused after scheduling, and how much of it can pruning reclaim?}

\myparagraph{Methodology}
The initial layout provides routing and magic-state resources before the scheduler knows which routes will be selected. This layout is intentionally conservative. After scheduling, Harvest traverses the routed instruction list and removes auxiliary resources that are never used by the final schedule. Data patches are preserved because they store logical qubits, but unused routing patches, magic-state patches, and ports can be removed.

\myparagraph{Analysis}
Fig.~\ref{fig:pacthes_removed_by_layout_line}(a) separates pruning effectiveness by layout template and resource type. The $\frac{2}{3}$-filling layout has the largest total reduction with 30.9\%, followed by $\frac{1}{9}$-filling with 26.5\% and $\frac{1}{4}$-filling with 24.2\%. Magic-state patches are reduced most aggressively, with reductions around 46.5--50.3\% depending on the layout. Routing-patch reductions are smaller but still meaningful, reaching 19.5\% for the $\frac{1}{9}$-filling layout.

Fig.~\ref{fig:pacthes_removed_by_layout_line}(b) shows the percentage of removed patches as circuit size changes. The pruner removes many patches in absolute terms for larger circuits because their initial layouts are larger. In relative terms, the largest reductions occur for smaller circuits, where the initial layout often contains more routing resources than the final schedule needs.

Fig.~\ref{fig:pacthes_removed_by_layout_line}(c) shows pruning by circuit family. Small and Feynman circuits leave the largest fraction of removable resources. The Small family removes 72.0\% of magic-state patches and 33.9\% of routing patches, while Feynman circuits remove 69.8\% of magic-state patches and 28.2\% of routing patches. In contrast, Clifford, QAOA, and QFT leave comparatively little unused routing space, indicating that these circuits use the layout more broadly or create stronger routing pressure.

\statementLeftAccent{Takeaway \#5}{green}{Pruning shrinks the initial layout by 24.2--30.9\% on average, removing about half of all magic-state patches, without altering the schedule.}

\section{Related Work}
\label{sec:related-work}
\myparagraph{General quantum compilation}
Other work targets gate-level qubit mapping and routing rather than fault-tolerant lattice surgery. Noise-adaptive heuristics place and route near-term circuits using hardware reliability data~\cite{murali2019noiseadaptive}, Amaro generates mapping-and-routing compilers automatically from a device model~\cite{molavi_2026_qubit_mapping}, and Paulihedral restructures Pauli-product circuits for block-wise synthesis and mapping~\cite{li2022paulihedral}. Classical qubit-mapping heuristics such as SABRE and its successors~\cite{original_sabre,ibm_sabre}, parallelized mapping~\cite{swin}, and multi-core routing~\cite{route_forcing_escofet} address the same NISQ-era problem. Harvest instead targets placement, routing, and scheduling for fault-tolerant lattice-surgery execution, where magic-state availability further constrains feasible schedules.

\myparagraph{Fault-tolerant quantum compilation}
Since optimizing lattice-surgery layouts under space-time costs is hard~\cite{Herr_2017}, systems rely on intermediate representations and heuristics. OpenSurgery and SurfBraid support representation, visualization, and resource estimation ~\cite{paler2020opensurgerytopologicalassemblies, paler2019surfbraidconcepttoolpreparing}, while the Lattice Surgery Compiler and \texttt{liblsqecc} provide complete compilation flows ~\cite{watkins2024high,leblond_realistic_2024}, and recent work extends lattice-surgery compilation beyond the surface code~\cite{herzog2025latticesurgerycompilationsurface}. Other systems address mapping and scheduling directly: DASCOT uses dependency-aware compilation and simulated annealing~\cite{molavi2025dascot}, Ecmas studies surface-code mapping and scheduling~\cite{zhu2023ecmasefficientcircuitmapping}, LaSsynth uses SAT-based subroutine synthesis~\cite{Tan_2024}, TopoLS combines ZX optimization with topology-aware compilation~\cite{zhou2026topolslatticesurgerycompilation}, and C-phase-aware compilation exploits commutativity under routing contention ~\cite{bharadwaj2026cphaseawarecompilationefficientfaulttolerant}. Harvest is complementary: it targets Pauli-product workloads and keeps routes, magic-state availability, placement, pruning, and scheduling explicit in H-IR.

\myparagraph{Architectural resource models}
Architectural approaches change the logical substrate or estimate costs at a higher level. LSQCA separates computation from dense scan-access memory ~\cite{kobori_lsqca_2025}, while the Azure Quantum Resource Estimator predicts costs under architectural and magic-state assumptions ~\cite{vandam2024usingazurequantumresource}. QuIRC co-designs a superconducting architecture with a routing card for lattice surgery~\cite{quirc}, and \texttt{tqec} provides tooling for topological error correction~\cite{tqec}. The closest system is Pure Magic scheduling~\cite{hofmeyr_scheduling_2026}, which studies scheduling with magic-state cultivation and dynamic ancilla-patch reuse. Harvest instead makes use of placement, scheduling, and pruning within a single pipeline.

\begin{table}[t]
\centering
\caption{Related-work comparison. \textit{Feature-level comparison of Harvest
with related lattice surgery compilation and architecture work. A check/ circle/ dash indicates full/ partial/ no support, respectively.}}
\label{tab:related-work-matrix}
\footnotesize
\renewcommand{\arraystretch}{1.15}

\begin{tabularx}{\linewidth}{
>{\raggedright\arraybackslash}X
*{4}{>{\centering\arraybackslash}p{1.1cm}}
}
\toprule
\textbf{Work} &
\textbf{Placement} &
\textbf{Scheduling} &
\textbf{Generality} &
\textbf{Pruning} \\
\midrule

DASCOT~\cite{molavi2025dascot} &
\fullsupport &
\fullsupport &
\partialsupport &
\nosupport \\

LS ~\cite{leblond_realistic_2024} &
\partialsupport &
\fullsupport &
\fullsupport &
\nosupport \\

Pure Magic~\cite{hofmeyr_scheduling_2026} &
\partialsupport &
\fullsupport &
\partialsupport &
\nosupport \\

LSQCA~\cite{kobori_lsqca_2025} &
\fullsupport &
\partialsupport &
\partialsupport &
\nosupport \\

\textbf{Harvest} &
\fullsupport &
\fullsupport &
\fullsupport &
\fullsupport \\

\bottomrule
\end{tabularx}
\end{table}

\myparagraph{Magic-state generation and optimization}
Magic states enable universal fault-tolerant computation~\cite{Bravyi_2005}, but preparing them is expensive. Distillation protocols reduce this cost, including 15-to-1~\cite{Bravyi_2005}, 10-to-2~\cite{meier2012magicstatedistillationfourqubitcode}, Bravyi-Haah~\cite{Bravyi_2012}, and catalyzed \(\ket{CCZ}\) factories~\cite{Gidney_2019}. Other work lowers magic-state demand through $T$-gate optimization~\cite{albarghouthi2026lineartimetgateoptimizationrandom} or unitary synthesis~\cite{Hao_2026}, studies non-deterministic production~\cite{awasthi2026pricepayoffnondeterminismfault}, flag-based injection~\cite{Chamberland_2020}, experimental preparation~\cite{ye2023logicalmagicstatepreparation,kim2024magicstateinjectionibm}, and cultivation~\cite{gidney2024magicstatecultivationgrowing,hirano_efficient_2025, rosenfeld2025magicstatecultivationsuperconducting}. Harvest exposes these as configurable magic-state availability models.

\myparagraph{Summary}
Table~\ref{tab:related-work-matrix} makes the gap concrete: no prior system combines placement, resource-aware scheduling, protocol generality, and post-scheduling pruning. Harvest closes this gap by co-optimizing placement, routing, scheduling, and magic-state availability under a protocol-agnostic resource model rather than treating them separately.

\section{Conclusion}
\label{sec:conclusion}

Placement, routing, scheduling, and magic-state supply for lattice-surgery compilation could not be optimized in isolation, since independent operations could still contend for the same ports, routes, and magic-state terminals. We presented Harvest, a resource-aware compiler that co-optimized magic-state consumption with circuit-aware placement and congestion-aware routing under a protocol-agnostic resource model, then reclaimed unused layout footprint after scheduling. Across standard benchmark suites, Harvest achieved a 4.83$\times$ average speedup over sequential execution (up to 17.8$\times$), improved schedule length by up to 1.35$\times$ through circuit-aware placement, and reclaimed up to 72.0\% of unused magic-state patches and 33.9\% of unused routing patches. 

\myparagraph{Artifact} Harvest will be released publicly along with the evaluation artifact.

\section*{Acknowledgments}
This work is funded by the Bavarian State Ministry of Science and the Arts as part of the Munich Quantum Valley (MQV), grant number 6090181.

\bibliographystyle{IEEEtran}
\bibliography{sample-base}

@String{Computing = "Computing" }

@String{Computer = "{IEEE} Computer" }

@String{Springer = "Springer-Verlag" }

@ArtifactSoftware{R,
    title = {R: A Language and Environment for Statistical Computing},
    author = {{R Core Team}},
    organization = {R Foundation for Statistical Computing},
    address = {Vienna, Austria},
    year = {2019},
    url = {https://www.R-project.org/},
}

@misc{hirano_efficient_2025,
	title = {Efficient magic state cultivation with lattice surgery},
	url = {http://arxiv.org/abs/2510.24615},
	doi = {10.48550/arXiv.2510.24615},
	number = {{arXiv}:2510.24615},
	publisher = {{arXiv}},
	author = {Hirano, Yutaka and Toshio, Riki and Itogawa, Tomohiro and Fujii, Keisuke},
	urldate = {2026-01-23},
	date = {2025-10-28},
	langid = {english},
	eprinttype = {arxiv},
	eprint = {2510.24615 [quant-ph]},
}

@article{litinski_game_2019,
	title = {A Game of Surface Codes: Large-Scale Quantum Computing with Lattice Surgery},
	volume = {3},
	issn = {2521-327X},
	url = {http://arxiv.org/abs/1808.02892},
	doi = {10.22331/q-2019-03-05-128},
	shorttitle = {A Game of Surface Codes},
	pages = {128},
	journal = {Quantum},
	shortjournal = {Quantum},
	author = {Litinski, Daniel},
	urldate = {2026-01-23},
	date = {2019-03-05},
    year = {2019},
	langid = {english},
	eprinttype = {arxiv},
	eprint = {1808.02892 [quant-ph]},
}

@article{leblond_realistic_2024,
	title = {Realistic Cost to Execute Practical Quantum Circuits using Direct Clifford+T Lattice Surgery Compilation},
	volume = {5},
	issn = {2643-6809, 2643-6817},
	url = {http://arxiv.org/abs/2311.10686},
	doi = {10.1145/3689826},
	pages = {1--28},
	number = {4},
	journal = {{ACM} Transactions on Quantum Computing},
	shortjournal = {{ACM} Transactions on Quantum Computing},
	author = {{LeBlond}, Tyler and Dean, Christopher and Watkins, George and Bennink, Ryan S.},
	urldate = {2026-01-23},
	date = {2024-12-31},
    year = {2024},
	langid = {english},
	eprinttype = {arxiv},
	eprint = {2311.10686 [quant-ph]},
}

@inproceedings{kobori_lsqca_2025,
	title = {{LSQCA}: Resource-Efficient Load/Store Architecture for Limited-Scale Fault-Tolerant Quantum Computing},
	url = {http://arxiv.org/abs/2412.20486},
	doi = {10.1109/HPCA61900.2025.00033},
	shorttitle = {{LSQCA}},
	pages = {304--320},
	booktitle = {2025 {IEEE} International Symposium on High Performance Computer Architecture ({HPCA})},
	author = {Kobori, Takumi and Suzuki, Yasunari and Ueno, Yosuke and Tanimoto, Teruo and Todo, Synge and Tokunaga, Yuuki},
	urldate = {2026-01-23},
	date = {2025-03-01},
    year = {2025},
	langid = {english},
	eprinttype = {arxiv},
	eprint = {2412.20486 [quant-ph]},
}

@article{silva_multi-qubit_2024,
	title = {Multi-qubit Lattice Surgery Scheduling},
	volume = {310},
	issn = {1868-8969},
	url = {http://arxiv.org/abs/2405.17688},
	doi = {10.4230/LIPIcs.TQC.2024.1},
	pages = {1:1--1:22},
	journal = {{LIPIcs}, Volume 310, {TQC} 2024},
	author = {Silva, Allyson and Zhang, Xiangyi and Webb, Zak and Kramer, Mia and Yang, Chan Woo and Liu, Xiao and Lemieux, Jessica and Chen, Ka-Wai and Scherer, Artur and Ronagh, Pooya},
	urldate = {2026-01-23},
	year = {2024},
	langid = {english},
	eprinttype = {arxiv},
	eprint = {2405.17688 [quant-ph]},
}

@misc{hofmeyr_scheduling_2026,
	title = {Scheduling Lattice Surgery with Magic State Cultivation},
	url = {http://arxiv.org/abs/2512.06484},
	doi = {10.48550/arXiv.2512.06484},
	number = {{arXiv}:2512.06484},
	publisher = {{arXiv}},
	author = {Hofmeyr, Steven and Weiden, Mathias and Kalloor, Justin and Kubiatowicz, John and Iancu, Costin},
	urldate = {2026-01-23},
	date = {2026-01-15},
    year = {2025},
	langid = {english},
	eprinttype = {arxiv},
	eprint = {2512.06484 [quant-ph]},
}

@misc{zhou2026topolslatticesurgerycompilation,
      title={TopoLS: Lattice Surgery Compilation via Topological Program Transformations}, 
      author={Junyu Zhou and Yuhao Liu and Ethan Decker and Justin Kalloor and Mathias Weiden and Kean Chen and Costin Iancu and Gushu Li},
      year={2026},
      eprint={2601.23109},
      archivePrefix={arXiv},
      primaryClass={quant-ph},
      url={https://arxiv.org/abs/2601.23109}, 
}

@inproceedings{Tan_2024,
   title={A SAT Scalpel for Lattice Surgery: Representation and Synthesis of Subroutines for Surface-Code Fault-Tolerant Quantum Computing},
   url={http://dx.doi.org/10.1109/ISCA59077.2024.00032},
   DOI={10.1109/isca59077.2024.00032},
   booktitle={2024 ACM/IEEE 51st Annual International Symposium on Computer Architecture (ISCA)},
   publisher={IEEE},
   author={Tan, Daniel Bochen and Niu, Murphy Yuezhen and Gidney, Craig},
   year={2024},
   month=jun, pages={325–339} }

@misc{zhu2023ecmasefficientcircuitmapping,
      title={Ecmas: Efficient Circuit Mapping and Scheduling for Surface Code}, 
      author={Mingzheng Zhu and Hao Fu and Jun Wu and Chi Zhang and Wei Xie and Xiang-Yang Li},
      year={2023},
      eprint={2312.15254},
      archivePrefix={arXiv},
      primaryClass={quant-ph},
      url={https://arxiv.org/abs/2312.15254}, 
}

@article{molavi2025dascot,
title={Dependency-Aware Compilation for Surface Code Quantum Architectures},
author={Molavi, Abbas and Xu, Ang and Tannu, Swamit S. and Albarghouthi, Aws},
journal={Proceedings of the ACM on Programming Languages},
year={2025},
eprint={2311.18042}
}

@article{watkins2024high,
  title={A high performance compiler for very large scale surface code computations},
  author={Watkins, George and Nguyen, Hoang Minh and Watkins, Keelan and Pearce, Steven and Lau, Hoi-Kwan and Paler, Alexandru},
  journal={Quantum},
  volume={8},
  pages={1354},
  year={2024},
  publisher={Verein zur F{\"o}rderung des Open Access Publizierens in den Quantenwissenschaften}
}

@misc{paler2019surfbraidconcepttoolpreparing,
      title={SurfBraid: A concept tool for preparing and resource estimating quantum circuits protected by the surface code}, 
      author={Alexandru Paler},
      year={2019},
      eprint={1902.02417},
      archivePrefix={arXiv},
      primaryClass={quant-ph},
      url={https://arxiv.org/abs/1902.02417}, 
}

@misc{paler2020opensurgerytopologicalassemblies,
      title={OpenSurgery for Topological Assemblies}, 
      author={Alexandru Paler and Austin G. Fowler},
      year={2020},
      eprint={1906.07994},
      archivePrefix={arXiv},
      primaryClass={quant-ph},
      url={https://arxiv.org/abs/1906.07994}, 
}

@article{Herr_2017,
   title={Optimization of lattice surgery is NP-hard},
   volume={3},
   ISSN={2056-6387},
   url={http://dx.doi.org/10.1038/s41534-017-0035-1},
   DOI={10.1038/s41534-017-0035-1},
   number={1},
   journal={npj Quantum Information},
   publisher={Springer Science and Business Media LLC},
   author={Herr, Daniel and Nori, Franco and Devitt, Simon J.},
   year={2017},
   month=sep }

@misc{vandam2024usingazurequantumresource,
      title={Using Azure Quantum Resource Estimator for Assessing Performance of Fault Tolerant Quantum Computation}, 
      author={Wim van Dam and Mariia Mykhailova and Mathias Soeken},
      year={2024},
      eprint={2311.05801},
      archivePrefix={arXiv},
      primaryClass={quant-ph},
      url={https://arxiv.org/abs/2311.05801}, 
}

@misc{rosenfeld2025magicstatecultivationsuperconducting,
      title={Magic state cultivation on a superconducting quantum processor}, 
      author={Emma Rosenfeld and Craig Gidney and Gabrielle Roberts and Alexis Morvan and Nathan Lacroix and Dvir Kafri and Jeffrey Marshall and Ming Li and Volodymyr Sivak and Dmitry Abanin and Amira Abbas and Rajeev Acharya and Laleh Aghababaie Beni and Georg Aigeldinger and Ross Alcaraz and Sayra Alcaraz and Trond I. Andersen and Markus Ansmann and Frank Arute and Kunal Arya and Walt Askew and Nikita Astrakhantsev and Juan Atalaya and Ryan Babbush and Brian Ballard and Joseph C. Bardin and Hector Bates and Andreas Bengtsson and Majid Bigdeli Karimi and Alexander Bilmes and Simon Bilodeau and Felix Borjans and Jenna Bovaird and Dylan Bowers and Leon Brill and Peter Brooks and Michael Broughton and David A. Browne and Brett Buchea and Bob B. Buckley and Tim Burger and Brian Burkett and Nicholas Bushnell and Jamal Busnaina and Anthony Cabrera and Juan Campero and Hung-Shen Chang and Silas Chen and Zijun Chen and Ben Chiaro and Liang-Ying Chih and Agnetta Y. Cleland and Bryan Cochrane and Matt Cockrell and Josh Cogan and Paul Conner and Harold Cook and Rodrigo G. Cortiñas and William Courtney and Alexander L. Crook and Ben Curtin and Martin Damyanov and Sayan Das and Dripto M. Debroy and Sean Demura and Paul Donohoe and Ilya Drozdov and Andrew Dunsworth and Valerie Ehimhen and Alec Eickbusch and Aviv Moshe Elbag and Lior Ella and Mahmoud Elzouka and David Enriquez and Catherine Erickson and Lara Faoro and Vinicius S. Ferreira and Marcos Flores and Leslie Flores Burgos and Sam Fontes and Ebrahim Forati and Jeremiah Ford and Brooks Foxen and Masaya Fukami and Alan Wing Lun Fung and Lenny Fuste and Suhas Ganjam and Gonzalo Garcia and Christopher Garrick and Robert Gasca and Helge Gehring and Robert Geiger and Élie Genois and William Giang and Dar Gilboa and James E. Goeders and Edward C. Gonzales and Raja Gosula and Stijn J. de Graaf and Alejandro Grajales Dau and Dietrich Graumann and Joel Grebel and Alex Greene and Jonathan A. Gross and Jose Guerrero and Loïck Le Guevel and Tan Ha and Steve Habegger and Tanner Hadick and Ali Hadjikhani and Michael C. Hamilton and Monica Hansen and Matthew P. Harrigan and Sean D. Harrington and Jeanne Hartshorn and Stephen Heslin and Paula Heu and Oscar Higgott and Reno Hiltermann and Jeremy Hilton and Hsin-Yuan Huang and Mike Hucka and Christopher Hudspeth and Ashley Huff and William J. Huggins and Lev B. Ioffe and Evan Jeffrey and Shaun Jevons and Zhang Jiang and Xiaoxuan Jin and Chaitali Joshi and Pavol Juhas and Andreas Kabel and Hui Kang and Kiseo Kang and Amir H. Karamlou and Ryan Kaufman and Kostyantyn Kechedzhi and Tanuj Khattar and Mostafa Khezri and Seon Kim and Paul V. Klimov and Can M. Knaut and Bryce Kobrin and Alexander N. Korotkov and Fedor Kostritsa and John Mark Kreikebaum and Ryuho Kudo and Ben Kueffler and Arun Kumar and Vladislav D. Kurilovich and Vitali Kutsko and Tiano Lange-Dei and Brandon W. Langley and Pavel Laptev and Kim-Ming Lau and Emma Leavell and Justin Ledford and Joy Lee and Kenny Lee and Brian J. Lester and Wendy Leung and Lily Li and Wing Yan Li and Alexander T. Lill and William P. Livingston and Matthew T. Lloyd and Aditya Locharla and Laura De Lorenzo and Erik Lucero and Daniel Lundahl and Aaron Lunt and Sid Madhuk and Aniket Maiti and Ashley Maloney and Salvatore Mandrà and Leigh S. Martin and Orion Martin and Eric Mascot and Paul Masih Das and Dmitri Maslov and Melvin Mathews and Cameron Maxfield and Jarrod R. McClean and Matt McEwen and Seneca Meeks and Anthony Megrant and Kevin C. Miao and Zlatko K. Minev and Reza Molavi and Sebastian Molina and Shirin Montazeri and Charles Neill and Michael Newman and Anthony Nguyen and Murray Nguyen and Chia-Hung Ni and Murphy Yuezhen Niu and Nicholas Noll and Logan Oas and William D. Oliver and Raymond Orosco and Kristoffer Ottosson and Alice Pagano and Agustin Di Paolo and Sherman Peek and David Peterson and Alex Pizzuto and Elias Portoles and Rebecca Potter and Orion Pritchard and Michael Qian and Chris Quintana and Ganesh Ramachandran and Arpit Ranadive and Matthew J. Reagor and Rachel Resnick and David M. Rhodes and Daniel Riley and Roberto Rodriguez and Emma Ropes and Lucia B. De Rose and Eliott Rosenberg and Dario Rosenstock and Elizabeth Rossi and Pedram Roushan and David A. Rower and Robert Salazar and Kannan Sankaragomathi and Murat Can Sarihan and Max Schaefer and Sebastian Schroeder and Henry F. Schurkus and Aria Shahingohar and Michael J. Shearn and Aaron Shorter and Noah Shutty and Vladimir Shvarts and Spencer Small and W. Clarke Smith and David A. Sobel and Barrett Spells and Sofia Springer and George Sterling and Jordan Suchard and Aaron Szasz and Alexander Sztein and Madeline Taylor and Jothi Priyanka Thiruraman and Douglas Thor and Dogan Timucin and Eifu Tomita and Alfredo Torres and M. Mert Torunbalci and Hao Tran and Abeer Vaishnav and Justin Vargas and Sergey Vdovichev and Guifre Vidal and Benjamin Villalonga and Catherine Vollgraff Heidweiller and Meghan Voorhees and Steven Waltman and Jonathan Waltz and Shannon X. Wang and Danni Wang and Brayden Ware and James D. Watson and Yonghua Wei and Travis Weidel and Theodore White and Kristi Wong and Bryan W. K. Woo and Christopher J. Wood and Maddy Woodson and Cheng Xing and Z. Jamie Yao and Ping Yeh and Bicheng Ying and Juhwan Yoo and Noureldin Yosri and Elliot Young and Grayson Young and Adam Zalcman and Ran Zhang and Yaxing Zhang and Ningfeng Zhu and Nicholas Zobrist and Zhenjie Zou and Hartmut Neven and Sergio Boixo and Cody Jones and Julian Kelly and Alexandre Bourassa and Kevin J. Satzinger},
      year={2025},
      eprint={2512.13908},
      archivePrefix={arXiv},
      primaryClass={quant-ph},
      url={https://arxiv.org/abs/2512.13908}, 
}

@article{Fowler_2012,
   title={Surface codes: Towards practical large-scale quantum computation},
   volume={86},
   ISSN={1094-1622},
   url={http://dx.doi.org/10.1103/PhysRevA.86.032324},
   DOI={10.1103/physreva.86.032324},
   number={3},
   journal={Physical Review A},
   publisher={American Physical Society (APS)},
   author={Fowler, Austin G. and Mariantoni, Matteo and Martinis, John M. and Cleland, Andrew N.},
   year={2012},
   month=sep }

@misc{gidney2024magicstatecultivationgrowing,
      title={Magic state cultivation: growing T states as cheap as CNOT gates}, 
      author={Craig Gidney and Noah Shutty and Cody Jones},
      year={2024},
      eprint={2409.17595},
      archivePrefix={arXiv},
      primaryClass={quant-ph},
      url={https://arxiv.org/abs/2409.17595}, 
}

@article{Bravyi_2005,
   title={Universal quantum computation with ideal Clifford gates and noisy ancillas},
   volume={71},
   ISSN={1094-1622},
   url={http://dx.doi.org/10.1103/PhysRevA.71.022316},
   DOI={10.1103/physreva.71.022316},
   number={2},
   journal={Physical Review A},
   publisher={American Physical Society (APS)},
   author={Bravyi, Sergey and Kitaev, Alexei},
   year={2005},
   month=feb }

@misc{meier2012magicstatedistillationfourqubitcode,
      title={Magic-state distillation with the four-qubit code}, 
      author={Adam M. Meier and Bryan Eastin and Emanuel Knill},
      year={2012},
      eprint={1204.4221},
      archivePrefix={arXiv},
      primaryClass={quant-ph},
      url={https://arxiv.org/abs/1204.4221}, 
}

@article{Bravyi_2012,
   title={Magic-state distillation with low overhead},
   volume={86},
   ISSN={1094-1622},
   url={http://dx.doi.org/10.1103/PhysRevA.86.052329},
   DOI={10.1103/physreva.86.052329},
   number={5},
   journal={Physical Review A},
   publisher={American Physical Society (APS)},
   author={Bravyi, Sergey and Haah, Jeongwan},
   year={2012},
   month=nov }

@article{Gidney_2019,
   title={Efficient magic state factories with a catalyzed {$\lvert CCZ\rangle$} to {$2\lvert T\rangle$} transformation},
   volume={3},
   ISSN={2521-327X},
   url={http://dx.doi.org/10.22331/q-2019-04-30-135},
   DOI={10.22331/q-2019-04-30-135},
   journal={Quantum},
   publisher={Verein zur Forderung des Open Access Publizierens in den Quantenwissenschaften},
   author={Gidney, Craig and Fowler, Austin G.},
   year={2019},
   month=apr, pages={135} }

@article{Chamberland_2020,
   title={Very low overhead fault-tolerant magic state preparation using redundant ancilla encoding and flag qubits},
   volume={6},
   ISSN={2056-6387},
   url={http://dx.doi.org/10.1038/s41534-020-00319-5},
   DOI={10.1038/s41534-020-00319-5},
   number={1},
   journal={npj Quantum Information},
   publisher={Springer Science and Business Media LLC},
   author={Chamberland, Christopher and Noh, Kyungjoo},
   year={2020},
   month=oct }

@misc{ye2023logicalmagicstatepreparation,
      title={Logical Magic State Preparation with Fidelity Beyond the Distillation Threshold on a Superconducting Quantum Processor}, 
      author={Yangsen Ye and Tan He and He-Liang Huang and Zuolin Wei and Yiming Zhang and Youwei Zhao and Dachao Wu and Qingling Zhu and Huijie Guan and Sirui Cao and Fusheng Chen and Tung-Hsun Chung and Hui Deng and Daojin Fan and Ming Gong and Cheng Guo and Shaojun Guo and Lianchen Han and Na Li and Shaowei Li and Yuan Li and Futian Liang and Jin Lin and Haoran Qian and Hao Rong and Hong Su and Shiyu Wang and Yulin Wu and Yu Xu and Chong Ying and Jiale Yu and Chen Zha and Kaili Zhang and Yong-Heng Huo and Chao-Yang Lu and Cheng-Zhi Peng and Xiaobo Zhu and Jian-Wei Pan},
      year={2023},
      eprint={2305.15972},
      archivePrefix={arXiv},
      primaryClass={quant-ph},
      url={https://arxiv.org/abs/2305.15972}, 
}

@misc{kim2024magicstateinjectionibm,
      title={Magic State Injection on IBM Quantum Processors Above the Distillation Threshold}, 
      author={Younghun Kim and Martin Sevior and Muhammad Usman},
      year={2024},
      eprint={2412.01446},
      archivePrefix={arXiv},
      primaryClass={quant-ph},
      url={https://arxiv.org/abs/2412.01446}, 
}

@misc{awasthi2026pricepayoffnondeterminismfault,
      title={Price and Payoff: Non-Determinism in Fault Tolerant Quantum Computation}, 
      author={Aditi Awasthi and Sayam Sethi and Sahil Khan and Gokul Subramanian Ravi and Jonathan Mark Baker},
      year={2026},
      eprint={2605.07983},
      archivePrefix={arXiv},
      primaryClass={quant-ph},
      url={https://arxiv.org/abs/2605.07983}, 
}

@book{Nielsen_Chuang_2010, place={Cambridge}, title={Quantum Computation and Quantum Information: 10th Anniversary Edition}, publisher={Cambridge University Press}, author={Nielsen, Michael A. and Chuang, Isaac L.}, year={2010}}

@article{AaronsonGottesman2004, author={Aaronson, Scott and Gottesman, Daniel},
title={Improved Simulation of Stabilizer Circuits}, journal={arXiv preprint
quant-ph/0406196}, year={2004}, eprint={quant-ph/0406196},
archivePrefix={arXiv}}

@article{Shor1995, author={Shor, Peter W.}, title={Scheme for Reducing
Decoherence in Quantum Computer Memory}, journal={Physical Review A},
volume={52}, number={4}, pages={R2493--R2496}, year={1995}, doi={10.1103/
PhysRevA.52.R2493}}

@article{AharonovBenOr2008, author={Aharonov, Dorit and Ben-Or, Michael},
title={Fault-Tolerant Quantum Computation with Constant Error Rate},
journal={SIAM Journal on Computing}, volume={38}, number={4},
pages={1207--1282}, year={2008}, doi={10.1137/S0097539799359385}}

@article{Gottesman2009QECReview, author={Gottesman, Daniel}, title={An
Introduction to Quantum Error Correction and Fault-Tolerant Quantum
Computation}, journal={arXiv preprint arXiv:0904.2557}, year={2009},
eprint={0904.2557}, archivePrefix={arXiv}}

@article{DennisEtAl2002, author={Dennis, Eric and Kitaev, Alexei and Landahl,
Andrew and Preskill, John}, title={Topological Quantum Memory},
journal={Journal of Mathematical Physics}, volume={43}, number={9},
pages={4452--4505}, year={2002}, eprint={quant-ph/0110143},
archivePrefix={arXiv}}

@article{HorsmanEtAl2012, author={Horsman, Clare and Fowler, Austin G. and
Devitt, Simon and Van Meter, Rodney}, title={Surface Code Quantum Computing by
Lattice Surgery}, journal={New Journal of Physics}, volume={14}, number={12},
pages={123011}, year={2012}, doi={10.1088/1367-2630/14/12/123011}}

@misc{li2022qasmbenchlowlevelqasmbenchmark,
      title={QASMBench: A Low-level QASM Benchmark Suite for NISQ Evaluation and Simulation}, 
      author={Ang Li and Samuel Stein and Sriram Krishnamoorthy and James Ang},
      year={2022},
      eprint={2005.13018},
      archivePrefix={arXiv},
      primaryClass={quant-ph},
      url={https://arxiv.org/abs/2005.13018}, 
}

@misc{cross2017openquantumassemblylanguage,
      title={Open Quantum Assembly Language}, 
      author={Andrew W. Cross and Lev S. Bishop and John A. Smolin and Jay M. Gambetta},
      year={2017},
      eprint={1707.03429},
      archivePrefix={arXiv},
      primaryClass={quant-ph},
      url={https://arxiv.org/abs/1707.03429}, 
}

@inproceedings{Hao_2026,
   title={Reducing T Gates with Unitary Synthesis},
   url={http://dx.doi.org/10.1145/3779212.3790210},
   DOI={10.1145/3779212.3790210},
   booktitle={Proceedings of the 31st ACM International Conference on Architectural Support for Programming Languages and Operating Systems, Volume 2},
   publisher={ACM},
   author={Hao, Tianyi and Xu, Amanda and Tannu, Swamit},
   year={2026},
   month=Mar, pages={1589–1604} }

@misc{albarghouthi2026lineartimetgateoptimizationrandom,
      title={Linear-Time T-Gate Optimization via Random Abstraction}, 
      author={Aws Albarghouthi},
      year={2026},
      eprint={2605.13929},
      archivePrefix={arXiv},
      primaryClass={cs.PL},
      url={https://arxiv.org/abs/2605.13929}, 
}

@misc{bharadwaj2026cphaseawarecompilationefficientfaulttolerant,
      title={C-Phase-Aware Compilation for Efficient Fault-Tolerant Quantum Execution}, 
      author={Dhanvi Bharadwaj and Siddharth Dangwal and Yuewen Hou and Gokul Subramanian Ravi},
      year={2026},
      eprint={2605.14042},
      archivePrefix={arXiv},
      primaryClass={quant-ph},
      url={https://arxiv.org/abs/2605.14042}, 
}

@article{molavi_2026_qubit_mapping,
author = {Molavi, Abtin and Xu, Amanda and Cecchetti, Ethan and Tannu, Swamit and Albarghouthi, Aws},
title = {Generating Compilers for Qubit Mapping and Routing},
year = {2026},
issue_date = {January 2026},
publisher = {Association for Computing Machinery},
address = {New York, NY, USA},
volume = {10},
number = {POPL},
url = {https://doi.org/10.1145/3776720},
doi = {10.1145/3776720},
journal = {Proc. ACM Program. Lang.},
month = jan,
articleno = {78},
numpages = {30}
}

@inproceedings{murali2019noiseadaptive,
   title={Noise-Adaptive Compiler Mappings for Noisy Intermediate-Scale Quantum Computers},
   url={https://doi.org/10.1145/3297858.3304075},
   DOI={10.1145/3297858.3304075},
   booktitle={Proceedings of the Twenty-Fourth International Conference on Architectural Support for Programming Languages and Operating Systems},
   publisher={ACM},
   author={Murali, Prakash and Baker, Jonathan M. and Javadi-Abhari, Ali and Chong, Frederic T. and Martonosi, Margaret},
   year={2019},
   pages={1015--1029}
}

@inproceedings{li2022paulihedral,
   title={Paulihedral: A Generalized Block-Wise Compiler Optimization Framework For Quantum Simulation Kernels},
   url={https://doi.org/10.1145/3503222.3507715},
   DOI={10.1145/3503222.3507715},
   booktitle={Proceedings of the 27th ACM International Conference on Architectural Support for Programming Languages and Operating Systems},
   publisher={ACM},
   author={Li, Gushu and Wu, Anbang and Shi, Yunong and Javadi-Abhari, Ali and Ding, Yufei and Xie, Yuan},
   year={2022},
   pages={554--569}
}

@inproceedings{das2022afs,
   title={AFS: Accurate, Fast, and Scalable Error-Decoding for Fault-Tolerant Quantum Computers},
   url={https://doi.org/10.1109/HPCA53966.2022.00027},
   DOI={10.1109/HPCA53966.2022.00027},
   booktitle={2022 IEEE International Symposium on High-Performance Computer Architecture (HPCA)},
   publisher={IEEE},
   author={Das, Poulami and Pattison, Christopher A. and Manne, Srilatha and Carmean, Douglas M. and Svore, Krysta M. and Qureshi, Moinuddin K. and Delfosse, Nicolas},
   year={2022},
   pages={259--273}
}

@inproceedings{original_sabre,
    author = {Li, Gushu and Ding, Yufei and Xie, Yuan},
    title = {Tackling the Qubit Mapping Problem for NISQ-Era Quantum Devices},
    year = {2019},
    publisher = {Association for Computing Machinery},
    address = {New York, NY, USA},
    url = {https://doi.org/10.1145/3297858.3304023},
    doi = {10.1145/3297858.3304023},
    booktitle = {Proceedings of the Twenty-Fourth International Conference on Architectural Support for Programming Languages and Operating Systems},
    pages = {1001--1014},
    series = {ASPLOS '19}
}

@misc{ibm_sabre,
      title={LightSABRE: A Lightweight and Enhanced SABRE Algorithm},
      author={Henry Zou and Matthew Treinish and Kevin Hartman and Alexander Ivrii and Jake Lishman},
      year={2024},
      eprint={2409.08368},
      archivePrefix={arXiv},
      primaryClass={quant-ph},
      url={https://doi.org/10.48550/arXiv.2409.08368},
}

@article{swin,
    author = {Fu, Hao and Zhu, Mingzheng and Chen, Fangzheng and Zhang, Chi and Wu, Jun and Xie, Wei and Li, Xiang-Yang},
    title = {Effective and Efficient Parallel Qubit Mapper},
    year = {2025},
    publisher = {IEEE Press},
    volume = {44},
    number = {5},
    url = {https://doi.org/10.1109/TCAD.2024.3500784},
    doi = {10.1109/TCAD.2024.3500784},
    journal = {IEEE Transactions on Computer-Aided Design of Integrated Circuits and Systems},
    month = may,
    pages = {1774--1787}
}

@inproceedings{route_forcing_escofet,
  author={Escofet, Pau and Gonzalvo, Alejandro and Alarc{\'o}n, Eduard and Almud{\'e}ver, Carmen G. and Abadal, Sergi},
  booktitle={2024 IEEE International Conference on Quantum Computing and Engineering (QCE)},
  title={Route-Forcing: Scalable Quantum Circuit Mapping for Scalable Quantum Computing Architectures},
  year={2024},
  volume={01},
  pages={909-920},
  doi={10.1109/QCE60285.2024.00110},
  url={https://doi.org/10.1109/QCE60285.2024.00110}
}

@misc{herzog2025latticesurgerycompilationsurface,
      title={Lattice Surgery Compilation Beyond the Surface Code},
      author={Laura S. Herzog and Lucas Berent and Aleksander Kubica and Robert Wille},
      year={2025},
      eprint={2504.10591},
      archivePrefix={arXiv},
      primaryClass={quant-ph},
      url={https://arxiv.org/abs/2504.10591},
}

@misc{quirc,
      title={Co-Designed Superconducting Architecture for Lattice Surgery of Surface Codes with Quantum Interface Routing Card},
      author={Charles Guinn and Samuel Stein and Esin Tureci and Guus Avis and Chenxu Liu and Stefan Krastanov and Andrew A. Houck and Ang Li},
      year={2023},
      eprint={2312.01246},
      archivePrefix={arXiv},
      primaryClass={quant-ph},
      url={https://arxiv.org/abs/2312.01246},
}

@article{tqec,
    doi = {10.21105/joss.09142},
    url = {https://doi.org/10.21105/joss.09142},
    year = {2026},
    publisher = {The Open Journal},
    volume = {11},
    number = {120},
    pages = {9142},
    author = {Suau, Adrien and Zhang, Yiming and Thakre, Purva and Zhao, Yilun and Dubey, Kabir and Bolanos, Jose A. and Schelpe, Arabella and Hao, Tianyi and Seitz, Philip and Guerreschi, Gian Giacomo and P{\'e}rez, {\'A}ngela Elisa {\'A}lvarez and Stahn, Reinhard and Lenssen, Jerome and Reid, Brendan and Fowler, Austin},
    title = {tqec: A Python package for topological quantum error correction},
    journal = {Journal of Open Source Software}
}

@article{Acharya2025,
author={Acharya, Rajeev and Aleiner, Igor and Andersen, Trond I. and Ansmann, Markus and Arute, Frank and {Google Quantum AI and Collaborators}},
title={Quantum error correction below the surface code threshold},
journal={Nature},
year={2025},
month={Feb},
volume={638},
number={8052},
pages={920-926},
issn={1476-4687},
doi={10.1038/s41586-024-08449-y},
url={https://doi.org/10.1038/s41586-024-08449-y}
}

@article{bombin2006topological,
   title={Topological Quantum Distillation},
   volume={97},
   url={https://doi.org/10.1103/PhysRevLett.97.180501},
   DOI={10.1103/physrevlett.97.180501},
   number={18},
   journal={Physical Review Letters},
   publisher={American Physical Society (APS)},
   author={Bombin, H. and Martin-Delgado, M. A.},
   year={2006},
   month=oct
}

@article{gottesman1998theory,
  title = {Theory of fault-tolerant quantum computation},
  author = {Gottesman, Daniel},
  journal = {Phys. Rev. A},
  volume = {57},
  issue = {1},
  pages = {127-137},
  year = {1998},
  month = {Jan},
  publisher = {American Physical Society},
  doi = {10.1103/PhysRevA.57.127},
  url = {https://doi.org/10.1103/PhysRevA.57.127}
}

@article{terhal2015quantum,
  title = {Quantum error correction for quantum memories},
  author = {Terhal, Barbara M.},
  journal = {Rev. Mod. Phys.},
  volume = {87},
  issue = {2},
  pages = {307--346},
  year = {2015},
  month = {Apr},
  publisher = {American Physical Society},
  doi = {10.1103/RevModPhys.87.307},
  url = {https://doi.org/10.1103/RevModPhys.87.307}
}

@misc{decoding_latency_influence,
      title={Impacts of Decoder Latency on Utility-Scale Quantum Computer Architectures},
      author={Abdullah Khalid and Allyson Silva and Gebremedhin A. Dagnew and Tom Dvir and Oded Wertheim and Motty Gruda and Xiangzhou Kong and Mia Kramer and Zak Webb and Artur Scherer and Masoud Mohseni and Yonatan Cohen and Pooya Ronagh},
      year={2025},
      eprint={2511.10633},
      archivePrefix={arXiv},
      primaryClass={quant-ph},
      url={https://arxiv.org/abs/2511.10633},
}

@article{pymatching,
    author = "Higgott, Oscar and Gidney, Craig",
    title = "{Sparse Blossom: correcting a million errors per core second with minimum-weight matching}",
    eprint = "2303.15933",
    url={https://doi.org/10.22331/q-2025-01-20-1600},
    archivePrefix = "arXiv",
    primaryClass = "quant-ph",
    doi = "10.22331/q-2025-01-20-1600",
    journal = "Quantum",
    volume = "9",
    pages = "1600",
    year = "2025"
}

@article{sliding_window,
  title = {Scalable Surface-Code Decoders with Parallelization in Time},
  author = {Tan, Xinyu and Zhang, Fang and Chao, Rui and Shi, Yaoyun and Chen, Jianxin},
  journal = {PRX Quantum},
  volume = {4},
  issue = {4},
  pages = {040344},
  year = {2023},
  month = {Dec},
  publisher = {American Physical Society},
  doi = {10.1103/PRXQuantum.4.040344},
  url = {https://doi.org/10.1103/PRXQuantum.4.040344}
}

@article{crossOpenQASM3Broader2022,
  title         = {{{OpenQASM}} 3: {{A}} Broader and Deeper Quantum Assembly Language},
  author        = {Cross, Andrew W. and {Javadi-Abhari}, Ali and Alexander, Thomas and de Beaudrap, Niel and Bishop, Lev S. and Heidel, Steven and Ryan, Colm A. and Sivarajah, Prasahnt and Smolin, John and Gambetta, Jay M. and Johnson, Blake R.},
  year          = 2022,
  journal       = {ACM Transactions on Quantum Computing},
  volume        = {3},
  number        = {3},
  eprint        = {2104.14722},
  primaryclass  = {quant-ph},
  pages         = {1--50}
}

@misc{lattnerMLIRCompilerInfrastructure2020,
  title         = {MLIR: A Compiler Infrastructure for the End of Moore's Law},
  author        = {Lattner, Chris and Amini, Mehdi and Bondhugula, Uday and Cohen, Albert and Davis, Andy and Pienaar, Jacques and Riddle, River and Shpeisman, Tatiana and Vasilache, Nicolas and Zinenko, Oleksandr},
  year          = 2020,
  eprint        = {2002.11054},
  archivePrefix = {arXiv},
  primaryClass  = {cs.PL}
}

\end{document}